\documentclass[twocolumn,showpacs,showkeys,preprintnumbers]{revtex4}
\usepackage{graphicx}
\usepackage{dcolumn}
\usepackage{bm}

\begin{document}
\title{A proposed method for measurement of cosmic-ray chemical composition based on geomagnetic spectroscopy}
\author{Rajat K Dey}
\email{rkdey2007phy@rediffmail.com}
\author{Sandip Dam}
\email{sandip_dam@rediffmail.com}
\affiliation{Department of Physics, University of North Bengal, Siliguri, West Bengal, INDIA 734 013}

\begin{abstract}
The effect of the geomagnetic Lorentz force on the muon component of extensive air shower (EAS) has been studied in a Monte Carlo generated simulated data sample. This geomagnetic field affects the paths of muons in an EAS, causing a local contrast or polar asymmetry in the abundance of positive and negative muons about the shower axis. The asymmetry can be approximately expressed as a function of transverse separation between the positive and negative muons barycentric positions in the EAS through opposite quadrants across the shower core in the shower front plane. In the present study, it is found that the transverse muon barycenter separation and its maximum value obtained from the polar variation of the parameter are higher for Iron primaries than protons for highly inclined showers. Hence, in principle, these parameters can be exploited to the measurement of primary cosmic-ray chemical composition. Possibility of practical realization of the proposed method in a real experiment is briefly discussed.
\end{abstract}

\maketitle

\section{Introduction} 
Recent progress in astroparticle physics has improved our level of understanding of the outstanding unsolved problems concerning the origin, acceleration, and composition of primary cosmic rays (PCRs) during over past $103$ years since its discovery [1] at ultra-high energy (UHE) range with the continuous progress in experimental techniques and methods of measurements. Nowadays we have relatively more sensitive EAS experiments [2-5] consisting of a variety of modern detectors to observe the secondary components in an EAS that contribute important results. On the other hand, to arrive at any specific conclusions about cosmic-rays (CRs) from their indirect investigation it is very important to know how they interact with the atmosphere and how the EAS develops. This knowledge is obtained by Monte Carlo (MC) simulations which are tested against data. Hence air shower simulations are a crucial part of the design of air shower experiments and analysis of their data. But a MC technique relies heavily on high energy hadronic models which suffer to some degree of uncertainties from one model to another and increasing primary energy. Therefore, we are challenged to develop more accurate hadronic interaction models in place to predict robust results. Recently the LHC [6] data have been tuned in hadronic interaction models namely EPOS-LHC [7] and QGSJet-04 [8], those are now included in the MC code {\em CORSIKA} version 7.400 [9]. These recent efforts have improved the predictive power of EAS simulations significantly. 

To characterize an EAS initiated by any type of primary species, one has to know at least about its shower size (more specifically total charged particle size or electron size ($N_{e^{\pm}}$), muon size ($N_{{\mu}^{\pm}}$), or hadron size ($N_{h\bar{h}}$)) and shower age ($s$) [10]. Near shower maximum, the shower size/electron size is closely related to the energy of the primary particle [11]. On the other hand, EAS parameters like $N_{{\mu}^{\pm}}$, $s$ etc. have been used consistently to determine the nature of the shower initiating particle. Measurements of all these parameters of an individual shower are made either by an individual or hybrid detection method. These detection techniques require an array of scintillation detectors and assembly of proportional counters as muon tracking detectors (MTD) respectively. From the measurement, the particle density distribution data ($e^{\pm}$ or ${\mu}^{\pm}$) at the observation level are obtained. These data will then be used to reconstruct a shower with the help of a suitable lateral density profile. In the cascade theory, such a lateral density profile of cascade particles can be approximated at sea level by the well known Nishimura-Kamata-Greisen (NKG) structure function [12]. The radial distribution of cascade particles in an average EAS is generally assumed to be symmetrical in the plane perpendicular to the shower axis. But, presence of intrinsic fluctuations (due to stochastic nature of EAS development) from shower to shower, in addition higher zenith angles ($\Theta$) and geomagnetic effects (GEs) can perturb this axial symmetry noticeably. Such effects may even break the axial symmetry a little to the distribution of EAS particles even in vertically incident showers. Inclined showers though experience similar effects as vertical showers but manifest significantly large asymmetries. 

\begin{figure}
\begin{center}
\includegraphics[width=.5\textwidth,clip]{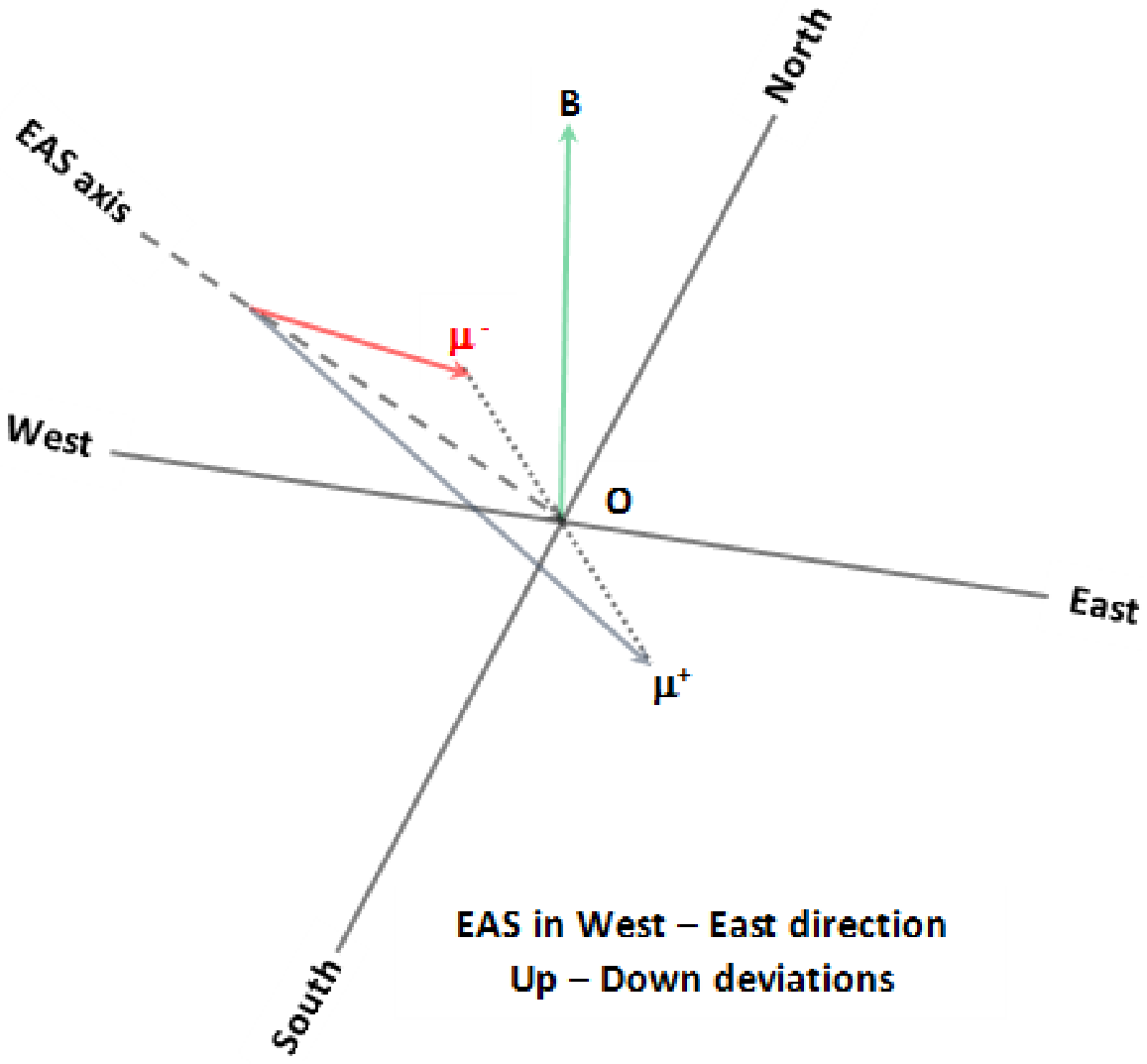}\hfill 
\includegraphics[width=.5\textwidth,clip]{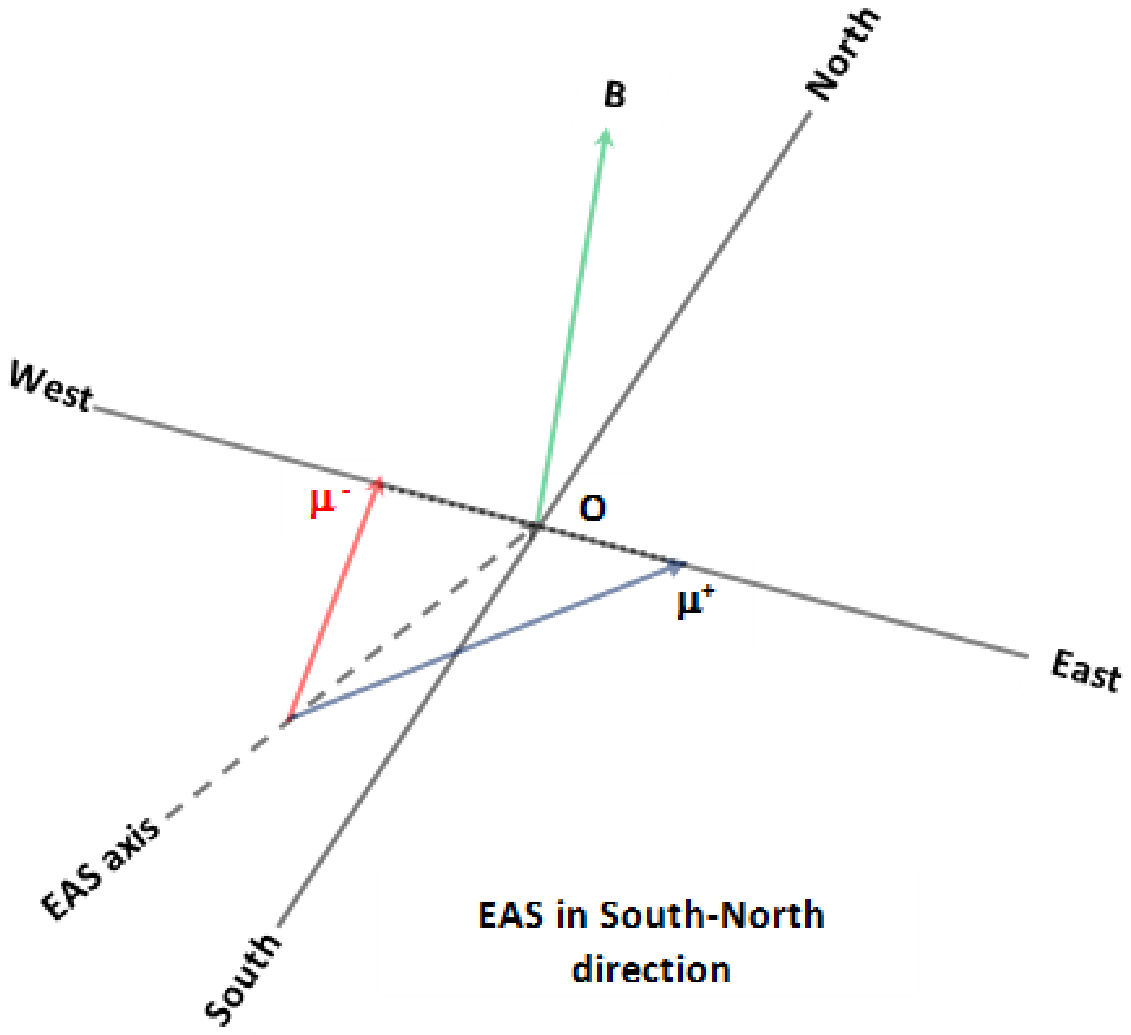}\hfill
\caption{The separation of $\mu^{+}$ and $\mu^{-}$ generated from a parent particle in an EAS by the geomagnetic field in two different situations.}
\end{center}
\end{figure}

Disregarding the GE, in ground array experiments the analysis of the density data is usually performed by assuming axial symmetry. Therefore, asymmetries would come from the polar variation of the charged EAS particles and unequal attenuation accounted from different locations of the EAS in the ground plane with inclined incidence. These are known as geometrical and atmospheric attenuation effects to polar asymmetries. To accentuate the GEs on the EAS charged particle distribution alone, the geometric and attenuation effects must be isolated or corrected out in the analysis. The data analysis technique which we are going to introduce here will remove the polar asymmetry caused by the geometric effect, while the asymmetry resulting from the attenuation effect would be ignored judiciously as muons come across very little attenuation in the atmosphere. In this work, it is shown that the asymmetry arises in polar distribution of $\mu$-s predominantly from the geomagnetic field (GF), may explore a new possibility for the determination of the chemical composition of primary CRs.
 
In this paper, we address the influence of the GF on the spatial distribution of $\mu$-s with a general limit on the $\Theta$ ($\Theta {\geq 50^{0}}$), valid for all applications of the technique, choosing the KASCADE experiment [2] site and look at the charge separation between $\mu^+$ and $\mu^-$ to arrive at a possible mass dependent parameter, called the {\it transverse muon barycenter separation} (TMBS). This TMBS is expected to be dependent on the nature of the primary particle and hence, in principle, the parameter can be exploited to estimate primary mass. The method presented here is applied to MC data simulated in three limited primary energy regions: $1-3$, $8-12$ and $98-102$ PeV (to obtain sufficient number of EAS events at these narrow energy ranges by our available computing power). The analysis described in this work is based on MC simulations carried out with the code {\em CORSIKA} ({\bf CO}smic {\bf R}ay {\bf SI}mulation for {\bf KA}scade) [9]. We also discuss the practical realization of the proposed method in a real experiment.

In this paper, basics of the influence of the GF on the EAS cascade, which are of direct relevance to this work, is discussed in the following section. In section III, we report on the simulation procedure adapted here. The MC data analysis technique is discussed in section IV. In section V, we present our results with discussion obtained from GEs on EAS muons. Finally, a possible experimental approach of the present method and concluding remarks are pointed out in section VI. 
                 
\section{Influence of the geomagnetic field on EAS cascade}
It is well known that the GF causes the East-West asymmetry on the primary CRs according to their rigidity and energy. The study of CRs with primary energies above $0.1$ PeV is usually based on the measurements of EASs, which are essentially cascades of secondary particles produced by interactions of CR particles with atmospheric nuclei. During the development of a CR cascade in the atmosphere, the GF affects the propagation of secondary charged particles in the shower: the perpendicular component of this field causes the trajectories of secondary charged particles to become curved, with positive and negative charged particles deflecting to produce a overall transverse barycenter separation. This aspect was first pointed out by Cocconi nearly sixty years back [13]. He further suggested that the geomagnetic broadening effect can be non-negligible in compare to the Coulomb scattering, particularly for the young showers. Since then, several studies have been carried out to address the influence of the GF on CR EAS and some important effects arising out of it were also reported (see for instance [14-30] and references therein). We should however state that some of these effects of the GF mentioned in the references have no direct relevance to this particular study.  
   
For the soft component ($e^{\pm}$), the radiation lengths in the atmosphere are short and they suffer many scatterings with $E_{e^{\pm}} < E_{cr.}$ (where, $E_{cr.} \approx 84$ MeV in air) and stronger bremsstrahlung effect (when $E_{e^{\pm}} > E_{cr.}$) thereby randomly changing the directions of their momenta relative to the GF. As a result, the lateral spread of electrons is mainly due to the multiple coulomb scattering and bremsstrahlung, and hence the effect of GF is less pronounced [31]. In contrast, after their generation from pion and kaon decays (mainly those in first few generations from their parent particles at great atmospheric heights), $\mu$-s travel much longer paths encountering negligible scattering with the medium (suffer lesser bremsstrahlung also), and hence come under the influence of GF noticeably. As a result, GE should be more pronounced in medium to high momenta $\mu$-s than lower ones, particularly for very large and strongly inclined showers. Such a situation is depicted by the figure 1, where, {\bf top}: separation of $\mu^{+}$ and $\mu^{-}$ generated from a parent particle in an EAS comes from the west and advances into the east direction, and {\bf bottom}: same as the {\bf top}, but the EAS comes from the south and advances into the north direction in the {\em CORSIKA} coordinate system. Besides, in highly inclined showers, a high percentage of EAS $e^{\pm}$ is absorbed in the atmosphere, may reduce shower-to-shower fluctuations to a  great extent. Exploiting this feature, some earlier MC simulation studies [24] reported that heavy nuclei and proton induced showers may be discriminated from the elliptic footprint of lateral muon distribution and the muon charge ratio (the ratio of $\mu^{+}$ to $\mu^{-}$ numbers) at convenient distances from the shower core. Through our preliminary study on this aspect of GF, their prediction was substantiated and reported in [32] from MC data. This paper includes more accurate data analysis technique, and quantifies different new features compared to earlier studies in [24,32], in order to obtain the primary mass information by the GF effects alone.   

\section{Simulation of EAS}
In the framework of the air shower simulation program {\em CORSIKA} of version $6.970$ [9], the EAS events are simulated by combining the high energy (above $80 {\rm GeV/n}$) hadronic interaction models QGSJet $01$ version $1$c [33] and EPOS $1.99$ [34], and the low energy (below $80 {\rm GeV/n}$) hadronic interaction model UrQMD [35]. The EGS$4$ program library is chosen for simulation of the electromagnetic (EM) component of shower that incorporates all the major interactions of electrons and photons [36]. We consider the US-standard atmospheric model with planar approximation, which works for $\Theta$ of the primary particles being less than $70^{\rm o}$ [37]. Events have also been generated at zenith angles ${70^{o}< \Theta <90^{o}}$ with CURVED  option of the atmosphere in the standard {\em CORSIKA} program [9]. The EAS events have been simulated at the geographical location of the experimental site of KASCADE (latitude $49.1^{o}$ N, longitude $8.4^{o}$ E, $110$ m a.s.l.) [2]. The GF with a homogeneous field approximation is considered. In order to examine the effect of the GF, EAS events are also simulated by switching off Earth's magnetic field. On the observation level, the detection kinetic energy thresholds are chosen as $3$ MeV for electrons ($e^{\pm}$) and $300$ MeV for muons. 

The EAS events have been generated for proton (p) and Iron (Fe) primaries at three different limited primary energy regions as mentioned in the Sect. I. At each energy range, we have taken five different narrow ranges of $\Theta$, $48^{o} - 52^{o}$, $53^{o} - 57^{o}$, $58^{o} - 62^{o} $, $63^{o} - 67^{o}$ and $68^{o} - 69^{o} $ respectively with FLAT option. A small sample of EAS events have also been simulated with ranges $73^{o} - 77^{o}$ and $78^{o} - 82^{o}$ with CURVED option of the atmosphere. In order to have at least some hints of fluctuations to our important observables, we have generated showers in those limited ranges of energy and $\Theta$. If we simulate EAS events in a wider range instead (say, $0 < \Theta < 70^0$), most of the events will then have $\Theta$ falling in $35^0 - 45^0$ range, but for which $\mu$-s will not remain a crucial component compare to other charged secondaries in the EAS. The proportion of $\mu$-s in an EAS rises with increasing $\Theta$ relative to the EM component, and hence the GF on $\mu$-s will be effective at higher $\Theta$ only. We have used thinning option of {\em CORSIKA} for the primary energy range $98 - 102$ PeV by taking $10^{-6}$ as thinning factor according to the optimum weight limitation [38]. We have restricted all showers to two azimuthal angles: $0^{o}$ and $90^{o}$ (only North and West directions in the coordinate system of {\em CORSIKA}). About $500$ showers have been generated for each $\Theta$ range in the energy region $1-3$ PeV and $100$ events for each $\Theta$ range in other two energy regions. 

\section{Data analysis method and selection criteria}

When the GF is disabled (setting $\sim 10^{-5}$ times smaller than the magnetic fields at the KASCADE location) in the simulation program, the lateral distribution of EAS charged particles possess cylindrical symmetry for all radial distances around the EAS axis in the shower front plane. In the observed plane, however, such cylindrical symmetry is distorted for inclined EAS due to the geometrical and the atmospheric attenuation effects. When the GF is switched on, polar asymmetry arising from the GE is superimposed with those caused by the geometric and attenuation effects. Therefore, for isolating or correcting out the effect contributed by the GF only from other distortions, such as those added by other two effects, the following part of this section has to be worked out during data analysis.    

\begin{figure}
\centering
\includegraphics[width=0.4\textwidth,clip]{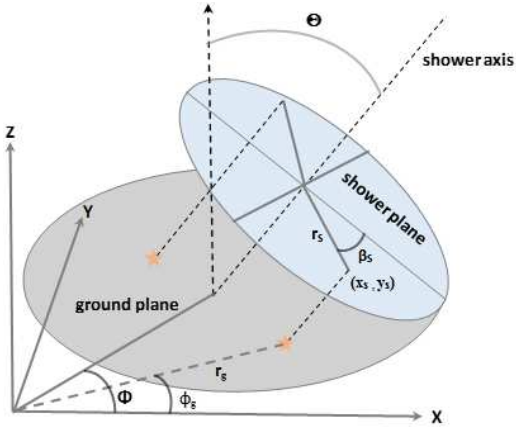} \hfill 
\caption{Sketch of the geometry of the ground plane and shower front plane used for the geometric correction in an inclined shower.}
\end{figure}

\begin{figure}
\centering
\includegraphics[width=0.4\textwidth,clip]{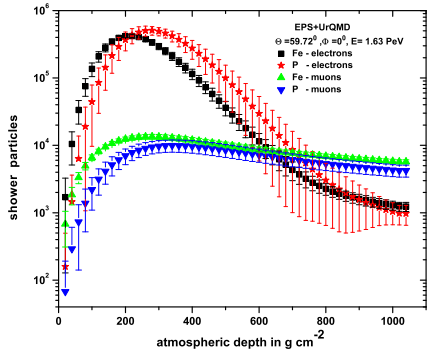} \hfill 
\caption{Average longitudinal distribution of electrons and muons for proton and iron induced showers.}
\end{figure}

Observables such as the muon density ($\rho_{\mu}$) or $N_{\mu}$ obtained from the analysis of simulated data are overestimated in the early region of an EAS while in the late region are underestimated. Hence, $\mu$-s in the first zone do undergo lesser attenuation than those arriving in the later zone. This development feature of the EAS accounts the attenuation contribution to the overall polar asymmetry in a particular observable under consideration, and is called the attenuation effect. The polar asymmetries are also present, if the given observable is measured in the ground plane. Such an asymmetry appears from another effect, is known as the geometric effect. To extract the actual variation introduced by the geometric and attenuation effects in EAS observables, as a first step it is necessary to take away the contribution added geometrically from data.

\begin{figure}
\centering
\includegraphics[width=0.4\textwidth,clip]{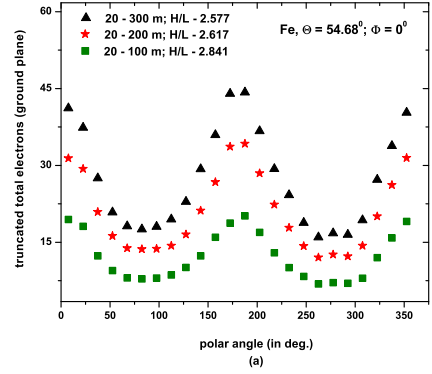} \hfill 
\includegraphics[width=0.4\textwidth,clip]{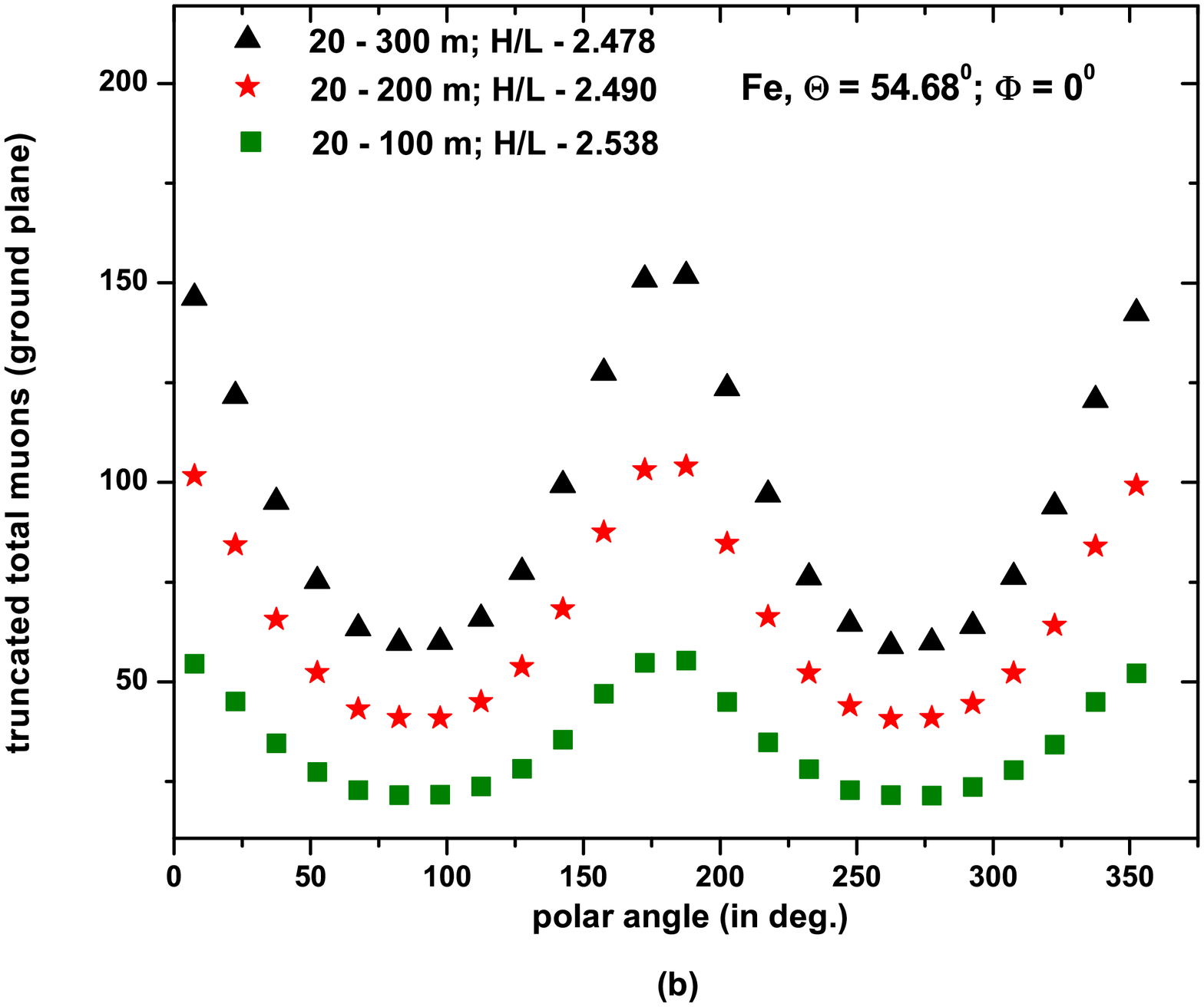} \hfill
\caption{Polar distribution of truncated electrons (Top) and muons (Bottom) from the simulated data in the ground plane.}
\end{figure} 

The correction results from the geometric effect can be briefly summarized by defining the relevant quantities involved. A projection procedure is applied for transforming the positional information of the point of impact of each particle in the ground plane onto the shower front plane. In the figure 2, we have shown the transformation of a point of impact by a cascade $\mu$ in the the ground ($\ast$ mark) with polar coordinates, ($r_{g}$,$\phi_{g}$) onto the shower plane with the set of coordinates, ($r_{s}$,$\beta_{s}$ or $x_s$, $y_s$). The necessary transformation relations are trivial and will be as follows,
  
\begin{equation}
r_{s} = r_{g} \sqrt{1-sin^{2}\Theta cos^{2}(\phi_{g}-\Phi).}
\end{equation} 

Using the figure 2, the corresponding Cartesian coordinates $P(x_{s}$,$y_{s})$ at the shower plane can be easily obtained as,
\begin{equation}
x_{s} = r_{g}cos(\phi_{g}-\Phi)cos\Theta
\end{equation} 
\begin{equation} 
y_{s} = r_{g}sin(\phi_{g}-\Phi)
\end{equation}

, where $\Phi$ denotes azimuthal angle of the EAS.

In the geometric correction, interaction due to attenuation process of $\mu$-s was not considered in the region between the two planes. The polar asymmetries in the shower front plane reflects the effect of attenuation of the $\mu$ component when the magnetic field is disabled, provided interaction of muons from the intermediate space between the two planes is taken into account. There are a number of reasons to ignore the attenuation effect on the $\mu$ component in the space between the two planes. All these reasons have been substantiated by the following studies using simulated data.

First, we have studied the average longitudinal development of $e^\pm$ and $\mu^\pm$ for higher $\Theta$, to see the attenuation power between the EM and muonic components of EAS after the shower maximum. The curves in the figure 3 illustrate how $\mu^\pm$ and $e^\pm$ contents of EAS vary during the EAS evolution from the depth of its size maximum. It is clear from the figure that the $\mu$ component remains nearly insulated from the atmospheric attenuation especially in the space between the two planes. This particular study involves all the charged muons with $E_{cut}^{\mu} \geq 300$ MeV, but the present study considers very energetic muons with $p_\mu \geq 100$ GeV/c. Important reasons behind such momenta selection will be discussed later. These energetic $\mu^\pm$, generated from their parent particles at great atmospheric heights are expected to be much weaker against attenuation. 

Secondly, polar asymmetries of truncated $e^\pm$ and $\mu^\pm$ contents in the ground plane normally arise due to both the geometric and attenuation effects when the GF is switched off in the simulation. But, a distinguishing character relating to the attenuation effect can be realized between these two polar variations for $e^\pm$ and $\mu^\pm$ numbers in the ground plane at $B \sim 10^{-5} \times B_{KAS}$. It is expected that the geometric effect will enhance the polar asymmetry equally to both $e^\pm$ and $\mu^\pm$ variations for a given truncated region within a polar angle range. The polar variations of $e^\pm$ and $\mu^\pm$ numbers in three different truncated regions ($20 - 100, 20 - 200, 20 - 300$) m within a polar angle interval of $\Delta \beta_g = 15^0$, each centered around increasingly different mean values of $\beta_g$ in the ground plane, are shown in plots through the figure 4 ($< 20$ m is not considered owing to large fluctuations near the EAS core). At a glance, these plots as if simply showing almost similar polar variations of $e^\pm$ and $\mu^\pm$ arising from these two above mentioned effects. But, some apparently unseen quantities as ratios of the highest-{\bf H} to the lowest-{\bf L} values of numbers of either types of particles separately from the figures 4a and 4b indicate that asymmetry in the EM component is about $4 - 12\%$ range higher than the $\mu$-s for three truncated limits in $r$. We have verified that these differences will increase further with increasing $\Theta$ and $E^{Thresh.}_{\mu}$. Hence, it is found that the attenuation contribution to the asymmetry by $\mu$-s is lower than the EM component from the comparison between these two figures, and may even be neglected in the space between these two planes under consideration. 

Moreover, some other studies related to the attenuation power of the $\mu$-component from the work in [39-40], backing the insignificant contribution by $\mu$-s in the concerned space between two planes according to the figure 2. The KASCADE-Grande studied the $N_{\mu}$ attenuation with the slant depth (sec$\Theta$), using the Constant Intensity Cut (CIC) method [41] in the primary energy range $10 - 10^3$ PeV, and they obtained little higher value for the attenuation length ($\Lambda_\mu$) of $N_{\mu}$ compared to MC data. The observed value of ($\Lambda_\mu$) was $\approx 1200$ g cm$^{-2}$, while the MC data yielded $\approx 900$ g cm$^{-2}$ and $\approx 980$ g cm$^{-2}$ respectively for lighter and relatively heavier compositions. At these energies, the attenuation length ($\Lambda_e$) of $N_e$ obtained by the KASCADE experiment was less than $\sim \frac{1}{5}$-th of the $\Lambda_\mu$ [39, 41-42]. This result also reiterates the fact that the attenuation of the $\mu$-component is much smaller than the EM component, and therefore has negligible effect on the polar asymmetry in the shower plane. A very recent study on the atmospheric attenuation of different EAS particles in the concerned space, introducing a newly defined  parameter, called 'a shift' of the EAS core also agrees with the insignificant trait of the attenuation effect from $\mu$-s [40]. 

\begin{table*}
\begin{center}
\begin{tabular}
{|l|l|l|l|l|r|} \hline
      Muons   & $r$-range (m)   & $67.5^0 - 112.5^0$   & $247.5^0 - 292.5^0$   & Total - $\mu$   & Q \\ 
&&&&& \\ \hline   
      $\mu^{+}$   & $30-60$   & $34.21\%$   & $6.45\%$   & 1816   & N \\ 
&&&&& \\ \hline
      $\mu^{-}$   & $30-60$   & $6.46\%$   & $33.55\%$   & 1805   & N \\ 
&&&&& \\ \hline
      $\mu^{+}$   & ${\bf 60-90}$   & ${\bf 37.54}\%$   & ${\bf 5.53}\%$   & ${\bf 1615}$   & {\bf Y} \\ 
&&&&& \\ \hline
      $\mu^{-}$   & ${\bf 60-90}$   & ${\bf 5.17}\%$   & ${\bf 36.86}\%$   & ${\bf 1605}$   & {\bf Y} \\ 
&&&&& \\ \hline
      $\mu^{+}$   & $90-120$   & $40.67\%$   & $3.87\%$   & 1319    & N \\ 
&&&&& \\ \hline
      $\mu^{-}$   & $90-120$   & $4.40\%$   & $40.68\%$   & 1273    & N \\ 
&&&&& \\ \hline       
\end{tabular}
\caption {Analysis showing an implementation of the selection of best possible muon detection regions in opposite sides keeping several factors in mind. Here, we have used Fe showers with $E = 98 - 102$ PeV, $\Theta = 68^0 - 69^0$ and $\Phi = 0^0$. The selection is made for charged muons with $p_{\mu} = 10^2 - 10^3$ GeV/c. Highlighted figures correspond qualified (Q) data from the selection.}  
\end{center}
\end{table*}

\begin{table*}
\begin{center}
\begin{tabular}
{|l|l||l|l|l||l|l|r|} \hline
      Muons    & $E$ (GeV)   & $p_\mu$ (GeV/c)   & $67.5^0 - 112.5^0$   & $247.5^0 - 292.5^0$    & $p_\mu$ (GeV/c)   & $67.5^0 - 112.5^0$   & $247.5^0 - 292.5^0$ \\ 
&&&&&&& \\ \hline   
      $\mu^{+}$   & $10^6$   & $1 - 10^2$   & $23.24\%$   & $11.45\%$   & ${\bf 10^2 - 10^3}$   & ${\bf 43.39}\%$   & $4.28\%$ \\ 
&&&&&&& \\ \hline
      $\mu^{-}$   & $10^6$   & $1 - 10^2$   & $12.02\%$   & $24.36\%$   & ${\bf 10^2 - 10^3}$   & $4.17\%$   & ${\bf 42.55}\%$ \\  
&&&&&&& \\ \hline
      $\mu^{+}$   & $10^7$   & $1 - 100$    & $22.71\%$   & $10.76\%$   & ${\bf 10^2 - 10^3}$   & ${\bf 35.79}\%$   & $5.94\%$ \\
&&&&&&& \\ \hline
      $\mu^{-}$   & $10^7$   & $1 - 10^2$    & $10.69\%$   & $23.51\%$   & ${\bf 10^2 - 10^3}$   & $5.07\%$   & ${\bf 39.11}\%$ \\
&&&&&&& \\ \hline
      $\mu^{+}$   & $10^8$   & $1 - 10^2$    & $24.62\%$   & $9.78\%$   & ${\bf 10^2 - 10^3}$   & ${\bf 37.54}\%$   & $5.53\%$ \\
&&&&&&& \\ \hline
      $\mu^{-}$   & $10^8$   & $1 - 10^2$    & $9.69\%$   & $25.52\%$   & ${\bf 10^2 - 10^3}$   & $5.18\%$   & ${\bf 36.86}\%$ \\ 
&&&&&&& \\ \hline       
\end{tabular}
\caption {Analysis showing an implementation of the selection of muons momenta in selected detection regions obtained from Table I. Here, we have used Fe showers with $E = 1-3$, $8-12$, and $98-102$ PeV, and $\Theta = 68^0 - 69^0$ and $\Phi = 0^0$. Highlighted figures correspond qualified data from the selection.} 
\end{center}
\end{table*} 

The analysis selects $\mu$-s which qualify two selection criteria simultaneously related to $\mu$ detection area and their energies or momenta. As already mentioned that this work considers relatively high energy $\mu$-s, those are expected to originate from first few interactions in the upper part of the atmosphere and travel much longer paths, and experiencing the GF for a longer time. This is consistent with the expectations since only the high momenta $\mu$-s will survive against attenuation especially in highly inclined showers. 

The main observables, TMBS and MTMBS (maximum value of TMBS) are expected to be dependent upon the energy of incoming $\mu$-s. This work has identified some suitable conditions for detection and measurement of these observables with reasonable $\mu$ numbers or densities. This situation essentially demands a best compromise among $\mu$ detector size, $\mu$ energy threshold and $N_{\mu}$. The visible effects of the GF are emphasized in the case of very inclined showers with high momentum $\mu$-s ($10^2 - 10^3$ GeV/c). The selection of high altitude site is useful for observing high energy $\mu$-s too. Otherwise, simulations of showers from very high energy primaries even at sea level, such as the KASCADE here, may produce high energy $\mu$-s as well.

The practical realization of the present approach in the work requires $\mu$ detectors at least at two opposite regions as shown in the last figure 12 (corresponding to showers coming from the North, West directions) in an array containing several scintillation detectors. Therefore, one of the main objectives of the work is to employ $\mu$ detectors into the array with some reasonable sizes from the point of view of their construction cost. We have used different combinations of momentum thresholds and $\mu$ detection areas as trials so that a better option may be surfaced, which will deliver an optimal asymmetry between $\mu^+$ and $\mu^-$ particles with a reasonable statistics. It has been seen that the low momenta $\mu$-s (below $10^2$ GeV/c) offer nearly symmetrical distribution in the X-Y shower plane whereas $\mu$-s  with momenta falling in the range $10^2 - 10^3$ GeV/c manifest a better polar asymmetry in $\mu^+$ and $\mu^-$ numbers in the annular region between $60$ m and $90$ m from the EAS core. Table I and Table II give a clear view on $\mu$-s selection explained above. The highlighted figures in both the tables correspond better combinations between $\mu$-s thresholds and their detection areas, for which an optimum local contrast in the abundance of $\mu^{+}$ and $\mu^{-}$ could be achieved. Data other than highlighted ones in those tables would either provide lower percentage of $\mu^{+}$ and $\mu^{-}$ abundances at two detector locations out of a relatively higher total or higher percentage of $\mu^{+}$ and $\mu^{-}$ abundances at detector locations out of a relatively lower total. Such combinations are found unsuitable for estimating TMBS and MTMBS with some anticipated uncertainties. In the upcoming sections, we shall use only these selected $\mu$-s having $p_{\mu} = 10^2 - 10^3$ GeV/c and $r = 60 - 90$ m for our important results.         

\section{Results and discussion}

\subsection{The polar asymmetries of the lateral distribution of EAS muons}

To examine the polar asymmetries of the lateral distributions of $\mu^{+}$ and $\mu^{-}$ due to the GF, we have estimated total number of each variety of particles over an arc region of truncated core distance range of $60 - 90$ m and central angle $15^{o}$ at different polar positions in both the planes. The polar variations of $\mu^{+}$ and $\mu^{-}$ ($p_{\mu} = 10^2 - 10^3$ GeV/c) in the shower plane with $B = B_{KAS.}$ and $B \sim 0$ are shown in figures 5a and 5b for Fe initiated showers corresponding to $ \Theta = 68^{o}$ and $\Phi = 0^0$. In the figure 5c, the polar asymmetries of these $\mu$-s in both the planes are depicted, while in the figure 5d, we have used showers having $\Phi = 90^0$ and the polar variation is limited to the shower plane only. However, it should be mentioned that these studies were also covered by other papers [25, 27] using $\mu$ muon densities from different perspectives.    

\begin{figure}
\centering
\includegraphics[width=0.45\textwidth,clip]{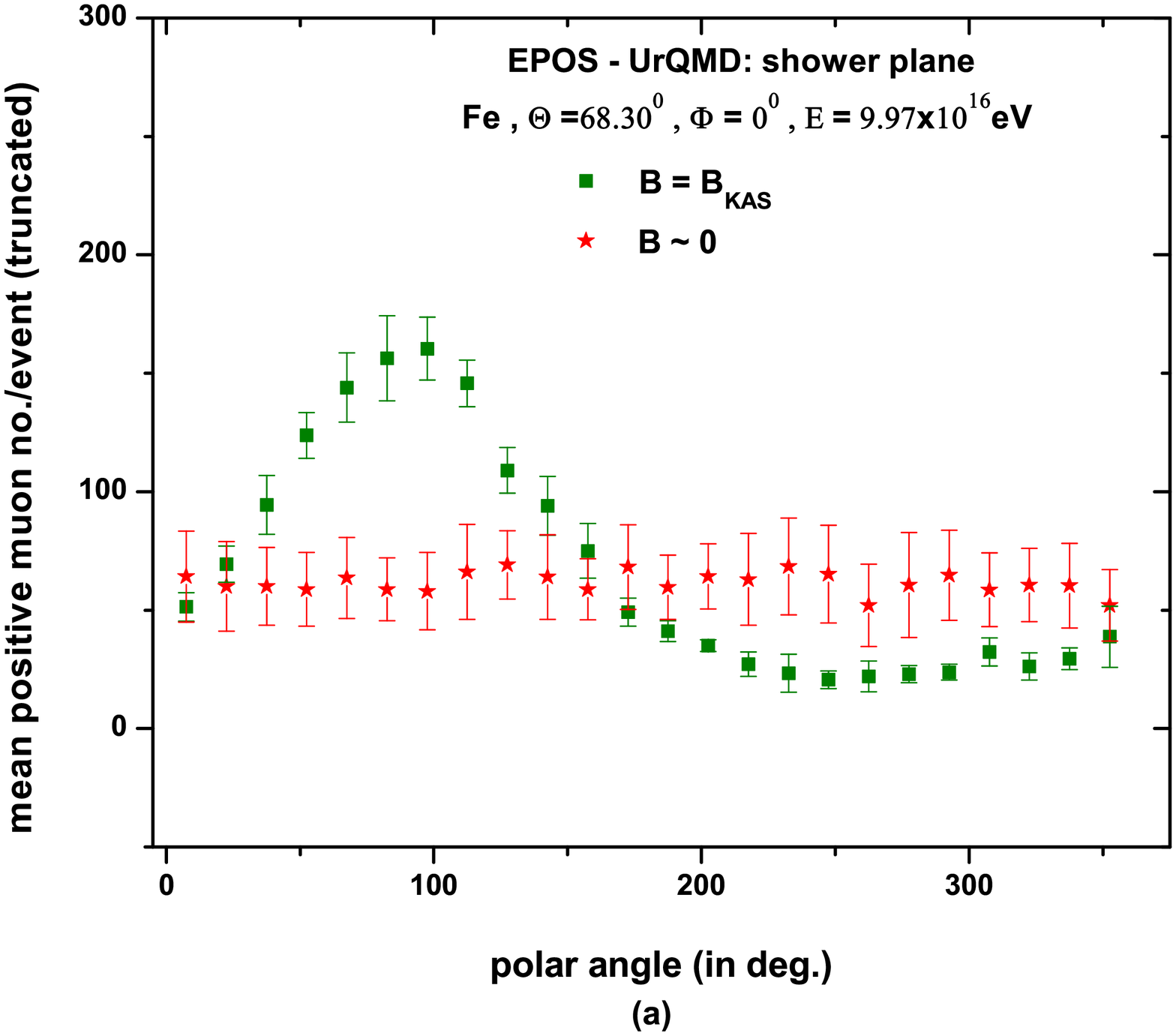} \hfill
\includegraphics[width=0.45\textwidth,clip]{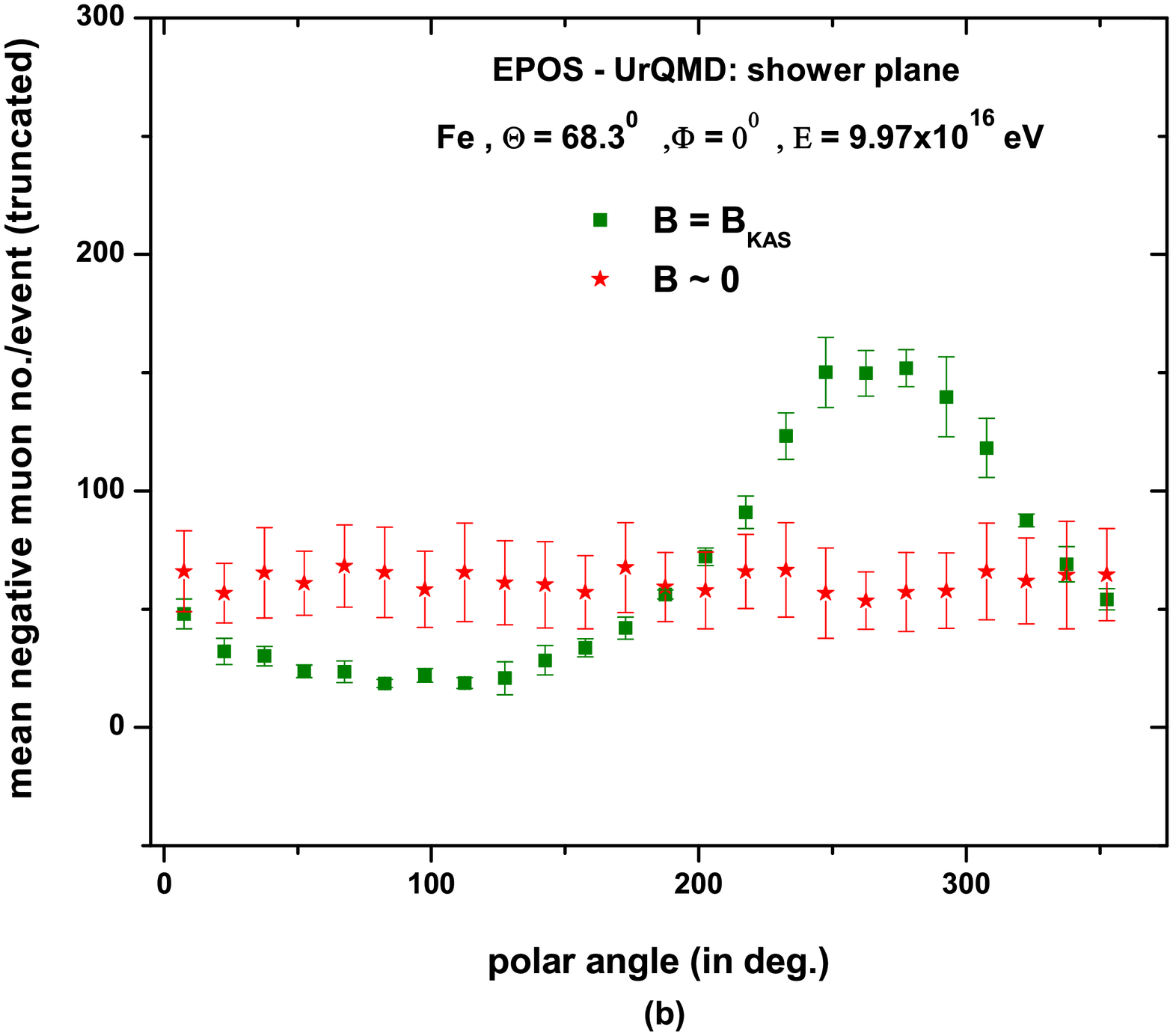} \hfill
\includegraphics[width=0.45\textwidth,clip]{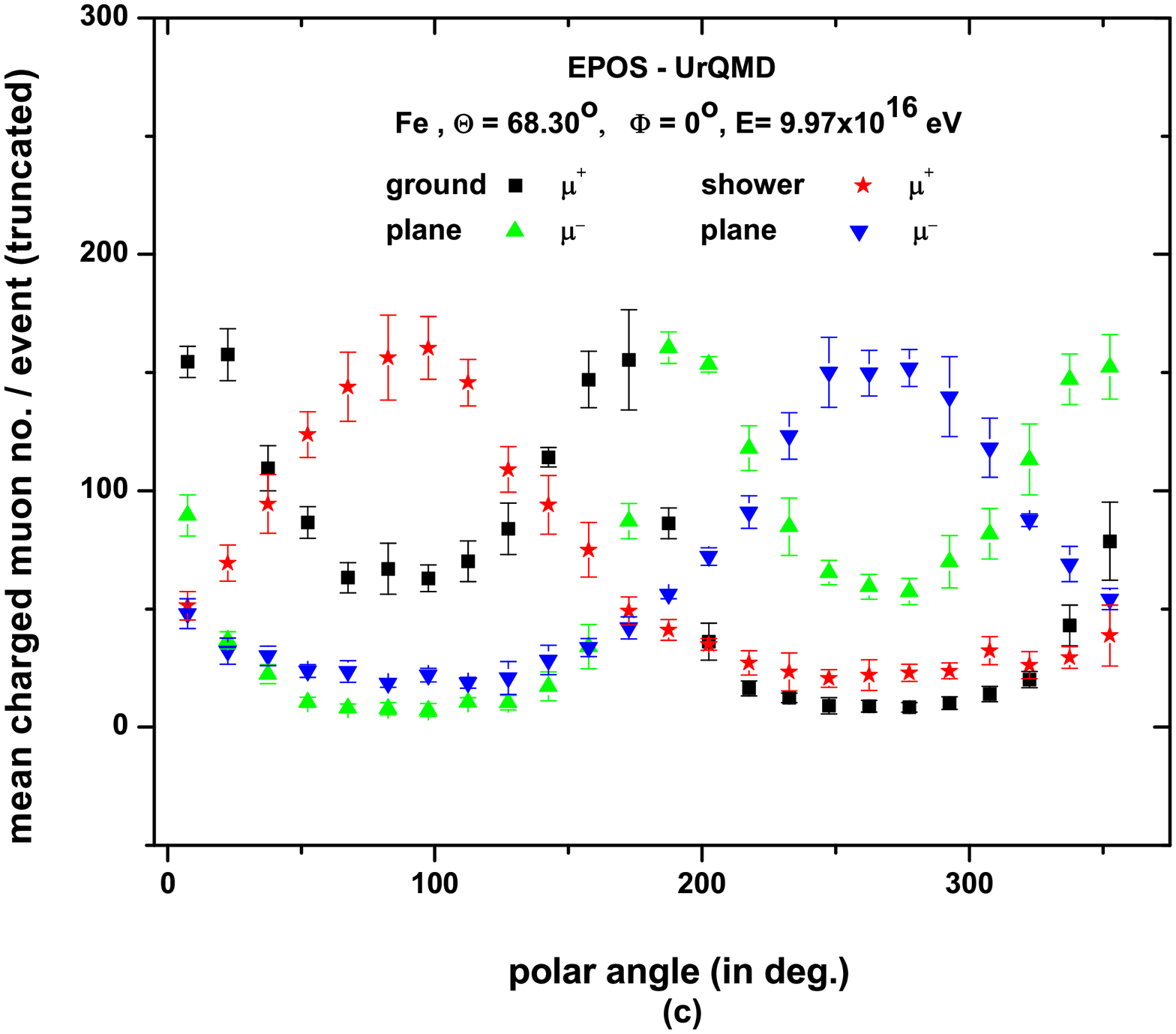} \hfill
\includegraphics[width=0.45\textwidth,clip]{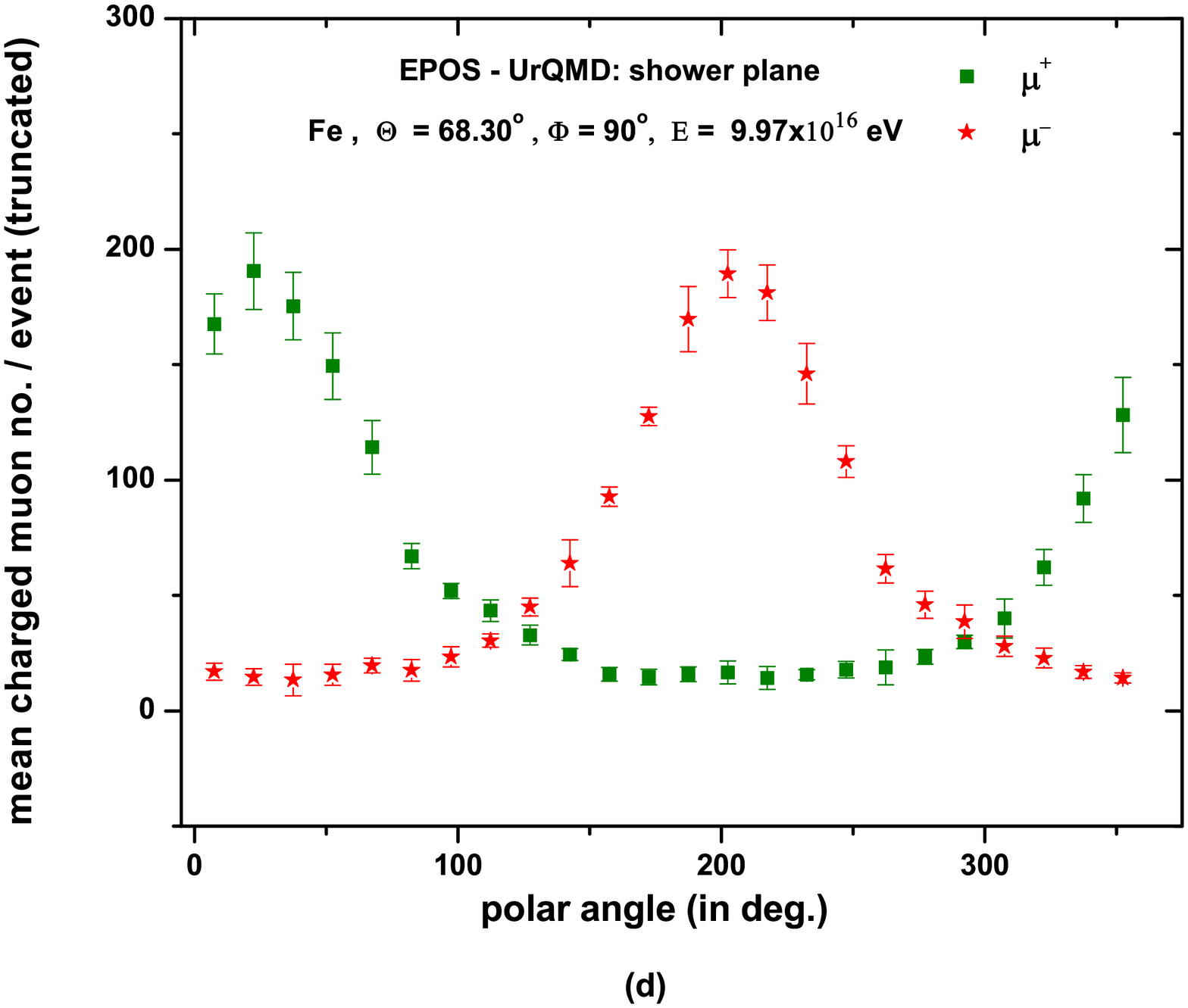} \hfill
\caption{The mean polar variation of $\mu^{+}$ and $\mu^{-}$ for iron primary arriving from the North and West directions.}
\end{figure}

For clear understanding of the influence of the GF, we have studied the polar variations of $\mu^{+}$ and $\mu^{-}$ contents separately in figures 5a and 5b by turning ON and OFF (dividing the components of the GF by a factor of $10^5$) the GF keeping all remaining parameters unchanged in the MC simulation program. The nature of these polar variations are as per our expectation. The figure 5c is a combined study on $\mu^{+}$ and $\mu^{-}$ variations including the ground plane for $B_{KAS.}$ only. Inclusion of the ground plane in the figure is merely for observing asymmetries compared to the shower plane. Our subsequent studies will not include the ground plane at all because it does not provide any additional information. The figure 5d represents the polar variations of $\mu^{+}$ and $\mu^{-}$ for showers coming from the west direction. It is noticed that the enhancements of $\mu^{+}$ and $\mu^{-}$ occur around $\beta_s \sim 97.5^0$ and $\sim 247.5^0$ when $\Phi = 0^o$ respectively. For $\Phi = 90^o$, such enhancements are found around $\beta_s \sim 112.5^0$ and $\sim 292.5^0$. These results would therefore validate our selection of polar angle ranges which have been worked out in tables I and II. Most of the figures cited above consider Fe initiated showers only, because of having little higher nuon contents (a generic feature) compared to p and other nuclei which may receive noticeable influence from the GF.           

\subsection{Primary mass sensitive parameters}
In the following, few relatively new EAS parameters related to the polar asymmetry of charged muons due to GF will be defined, and their sensitivity to the primary mass will be analyzed. These mass sensitive parameters are namely the TMBS, MTMBS and the eccentricity ($\epsilon$) of the elliptic lateral distribution of charged muons.
 
\subsubsection{The transverse muon barycenter separation and its maximum value}

To quantify the influence of GF as well as to identify some typical signatures of the nature of shower initiating primaries, we have estimated for each shower the coordinates of barycenters of $\mu^{+}$ and $\mu^{-}$ particles in the shower plane and thereby estimated TMBS, which actually measures the separation length between the barycenter positions of $\mu^{+}$ and $\mu^{-}$. For this purpose, we have introduced an operation that involves a rotation either clockwise or anti-clockwise sense of a hypothetical interior quadrant sector (IQS) for counting the $\mu^{+}$ and $\mu^{-}$ particles: the IQS is a region in the interior of a circle enclosed by a pair of arcs on opposite sides and a pair of diagonally aligned diameters on the other two sides making a central angle of $15^{o}$. However, the implemented IQS-S (special) in the work is designed by two isolated sectors from the IQS as defined above within the core distance range $60 - 90$ m along the diameters in opposite sides from the EAS core (a sketch is given - figure 6). 

\begin{figure}
\centering
\includegraphics[width=0.45\textwidth,clip]{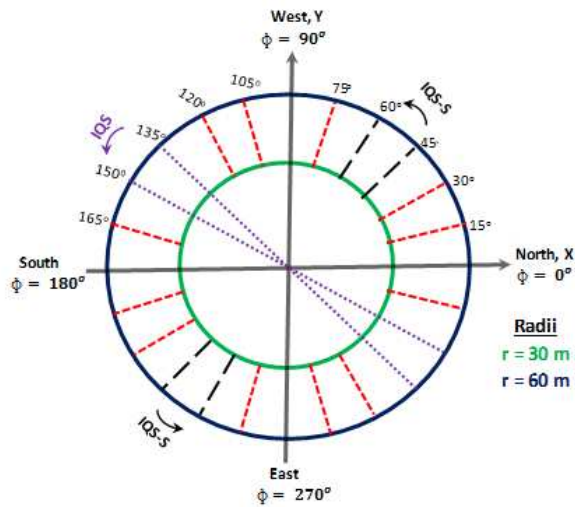} \hfill
\caption{Scanning of $\mu^{+}$ and $\mu^{-}$ particles by rotating IQS or IQS-S in anti-clockwise sense from $0^o$ to $180^o$.}
\end{figure}

The variation of the TMBS with IQS-S rotation for p and Fe initiated EASs falling in the $\Theta$-range $50^{o} - 80^{o}$ and arriving from North and West directions are studied using simulated data. In figures 7a and 7b, a comparison of such variations are shown for showers initiated by p and Fe primaries arriving from North and West directions at $\langle \Theta \rangle = 59.72^o$ and  $\langle E \rangle = 1.63$ PeV respectively. It is expected that $\mu^{+}$ particles experience GF greatly around polar angles $\sim 90^o$ while $\mu^{-}$ around $\sim 270^o$ (peaks in figures 5a and 5b) when $\Phi = 0^o$. The TMBS accordingly takes higher values corresponding to the orientation of the IQS-S through $\sim 90^o - 270^o$ in the shower plane which is reflected in the figure 7a. On the other hand, the GF effects on the $\mu$-component for showers coming from the West direction as the variation of the TMBS is exhibited in the figure 7b. In this case the parameter TMBS increases from the polar angle $0^o$ to $\sim 45^o$ and then decreases till around $105^o$, and again rises to a value that was obtained at $\sim 75^o$. It is clear that the TMBS is greater for Fe compared to p initiated showers in all the cases. In figures 8a - 8d, we have repeated the same study independently for p and Fe primaries for $\Phi = 0^o$ and $90^o$ where various curves in each of the figure 8 correspond different $\Theta $ of incidence of showers. As the $\Theta$ increases, $\mu$-s travel increased paths through the atmosphere receiving the GF effect for longer duration that induces more deflections to $\mu$-s and hence yield higher values for TMBS resulting from their greater deflections under the GF. The polar distribution of $\mu$ exhibits more and more asymmetry with increasing $\Theta$ and the feature is revealed from the gradual increase of TMBS in the neighborhood of polar angles $\sim 90^o$ and $\sim 45^o$ corresponding to $\Phi$ with $0^o$ and $90^o$ respectively irrespective of primaries.          

\begin{table*}
\begin{center}
\begin{tabular}
{|l|l|l|l|l|r|} \hline

      $\Theta$   & $50.23^{0}$   & $54.68^{0}$   & $59.72^{0}$   & $64.52^{0}$   & $68.3^{0}$ \\ 
&&&&&  \\ \hline

      Proton   & $0.246\pm 0.011$   & $0.422\pm 0.011$   & $0.677\pm 0.014$   & $0.706\pm 0.013$   & $0.799\pm 0.012$ \\ 
&&&&&  \\ \hline

     Iron   & $0.406\pm 0.011$   & $0.588\pm 0.014$   & $0.702\pm 0.014$   & $0.838\pm 0.012$   & $0.910\pm 0.009$ \\ 
&&&&&  \\ \hline

\end{tabular} 
\end{center}
\end{table*} 

\begin{table*}
\begin{center}
\begin{tabular}
{|l|l|l|l|l|r|} \hline

      $\Theta$   & $50.23^{0}$   & $54.68^{0}$   & $59.72^{0}$   & $64.52^{0}$   & $68.3^{0}$ \\ 
&&&&&  \\ \hline

      Proton   & $0.262\pm 0.005$   & $0.406\pm 0.012$   & $0.551\pm 0.013$   & $0.721\pm 0.013$   & $0.788\pm 0.012$ \\ 
&&&&&  \\ \hline

     Iron   & $0.427\pm 0.006$   & $0.596\pm 0.014$   & $0.708\pm 0.014$   & $0.792\pm 0.014$   & $0.913\pm 0.010$ \\ 
&&&&&  \\ \hline

\end{tabular}
\caption {The eccentricity parameters for showers initiated by proton and iron primaries and coming from the North direction: Top-QGSJet model; Bottom-EPOS model.} 
\end{center}
\end{table*} 

\begin{figure}
\centering
\includegraphics[width=0.45\textwidth,clip]{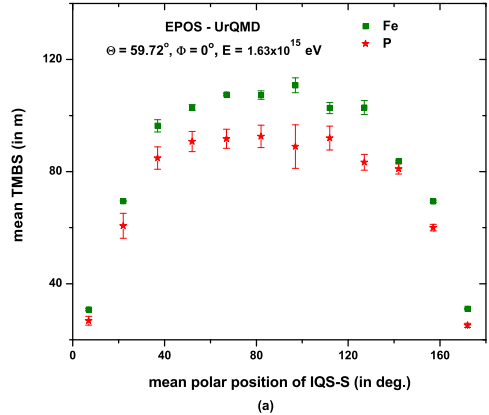} \hfill
\includegraphics[width=0.45\textwidth,clip]{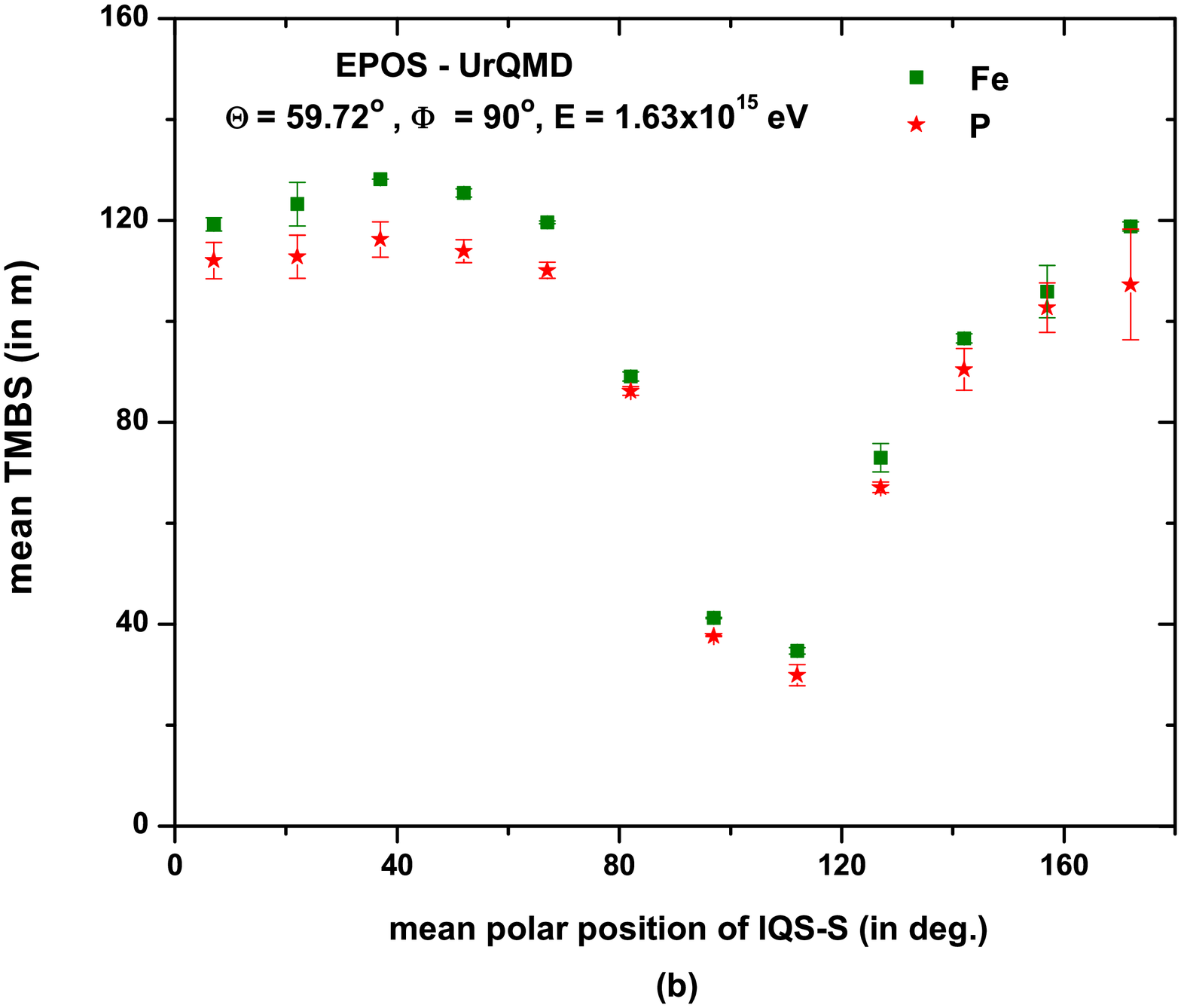} \hfill
\caption{Polar variation of the mean TMBS for p and Fe showers arriving from North (fig. a) and West directions (fig. b).}
\end{figure}
 
\begin{figure}
\centering
\includegraphics[width=0.45\textwidth,clip]{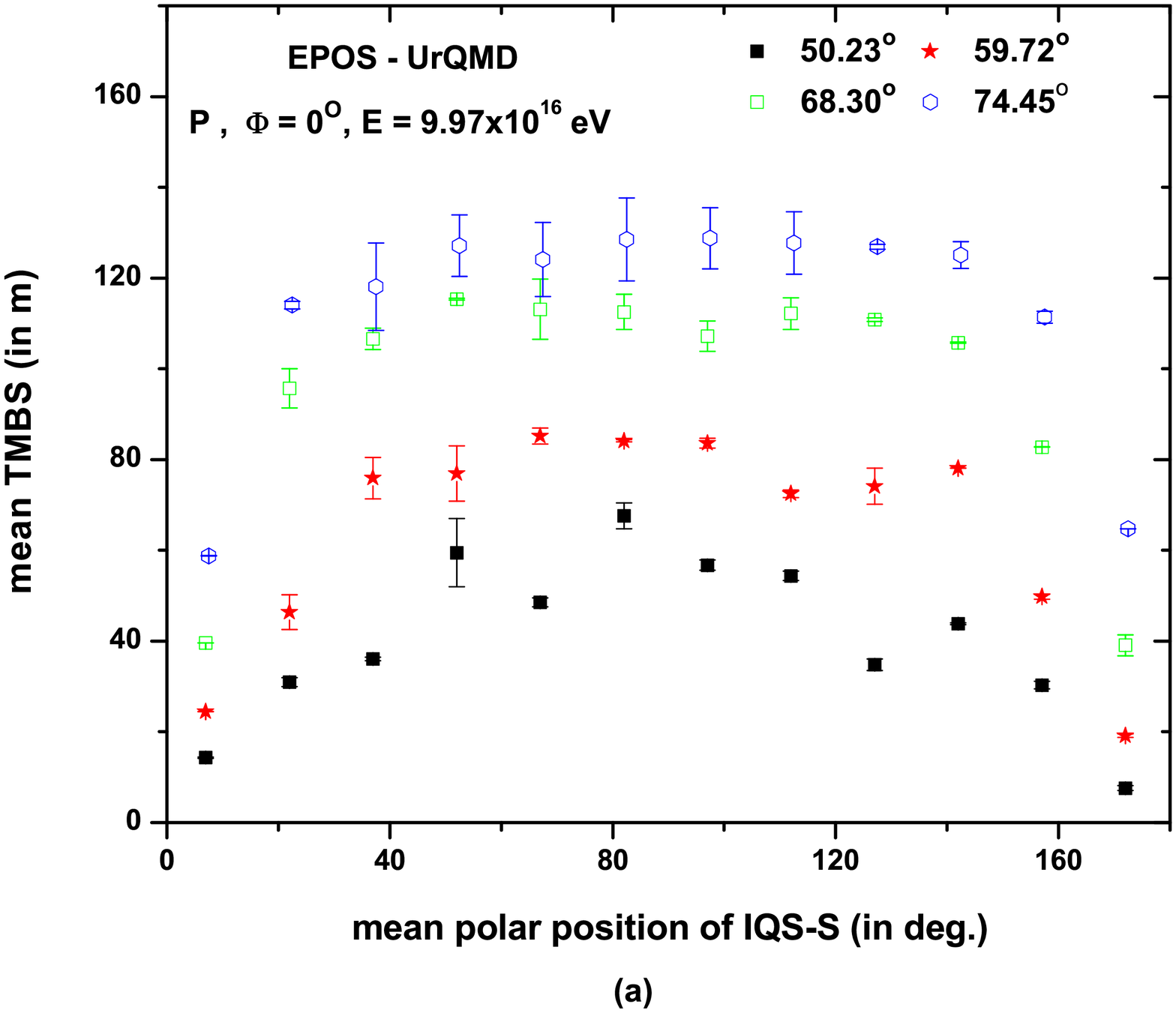} \hfill
\includegraphics[width=0.45\textwidth,clip]{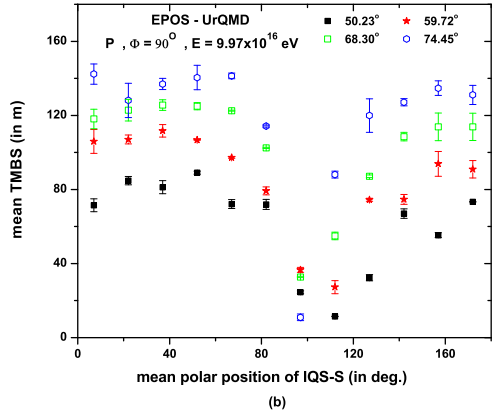} \hfill
\includegraphics[width=0.45\textwidth,clip]{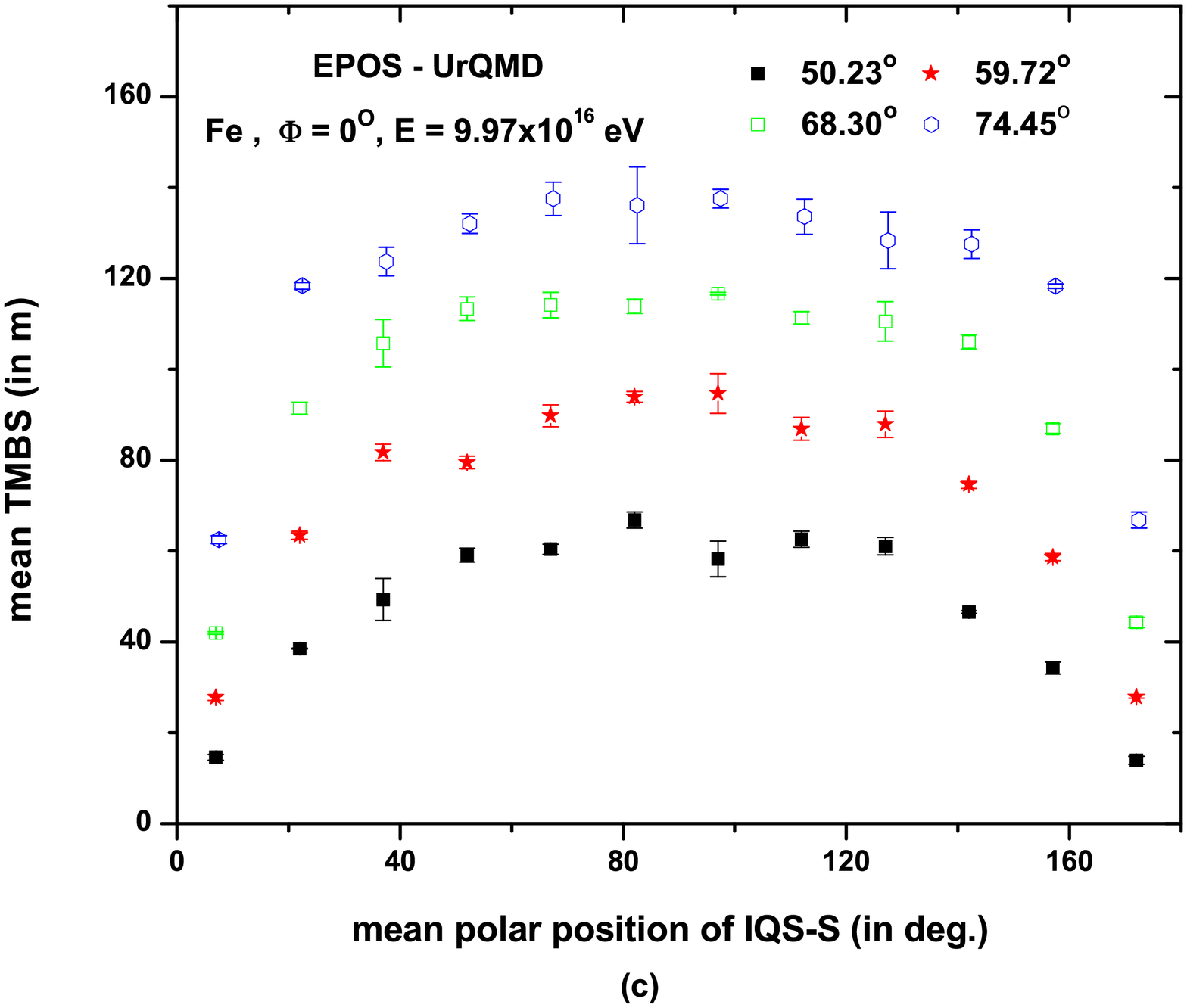} \hfill
\includegraphics[width=0.45\textwidth,clip]{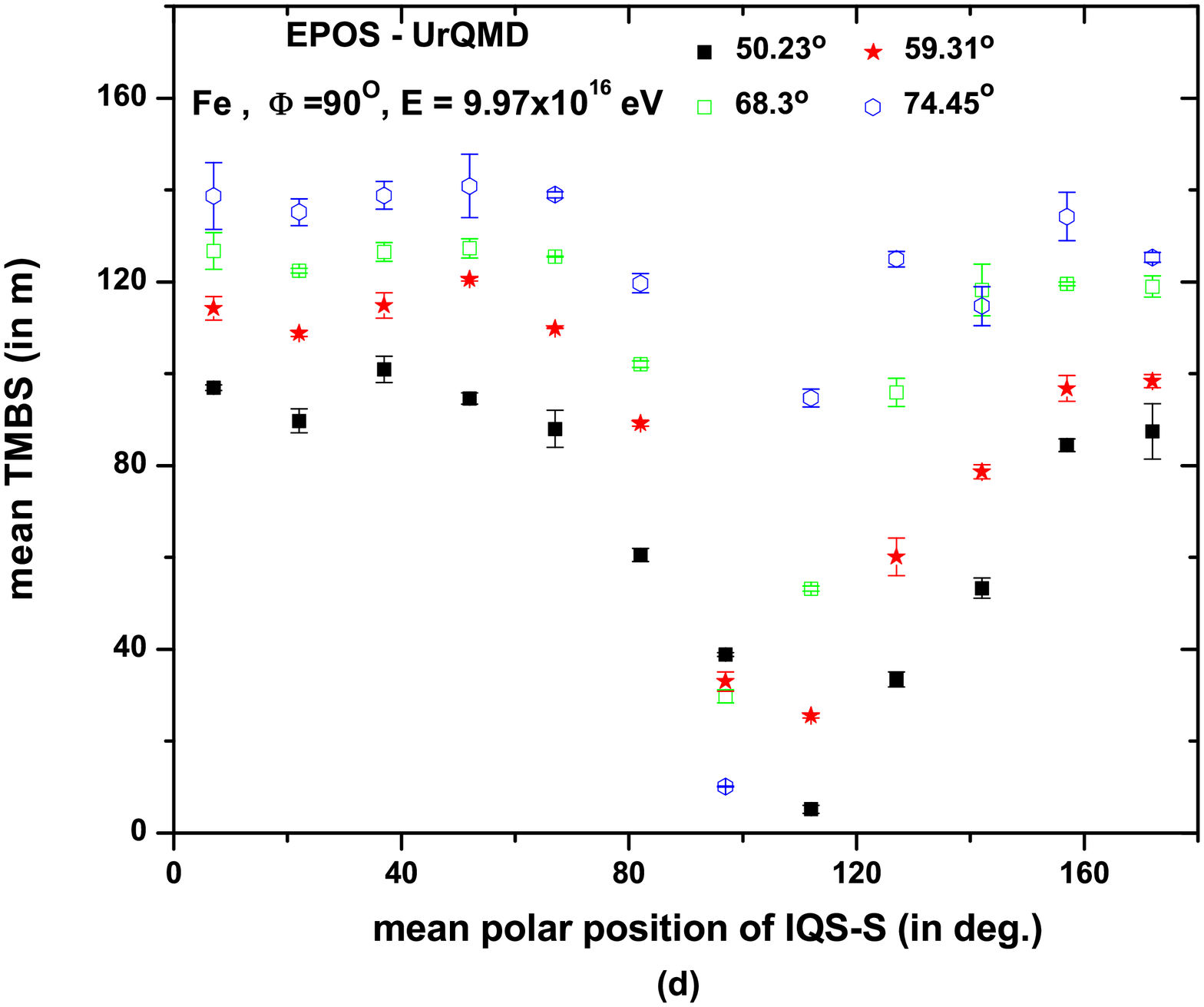} \hfill
\caption{Comparison of polar variations of TMBS for p and Fe showers arriving from the North and West directions with different zenith angles of incidence.}
\end{figure}

\begin{figure}
\centering
\includegraphics[width=0.45\textwidth,clip]{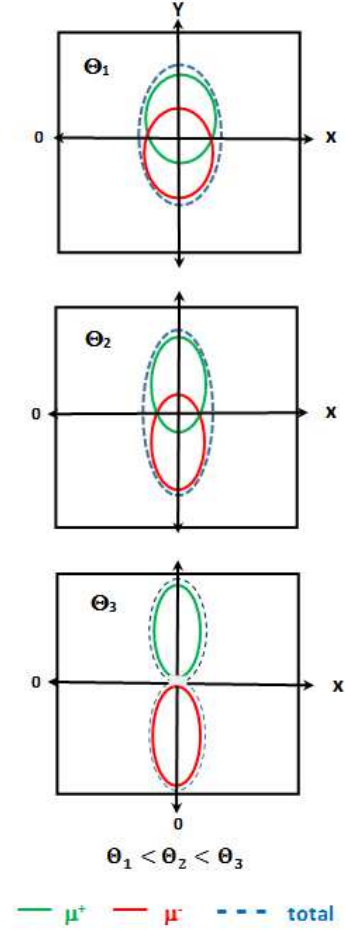} \hfill
\caption{Combined elliptic contour resulting from a pair of smaller iso-density contours of $\mu^+$ and $\mu^-$ with increasing $\Theta$ for $\Phi \sim 0^o$. At $\Theta_3$ the overall ellipse gets 8-shaped pattern.}
\end{figure}

\begin{figure}
\centering
\includegraphics[width=0.45\textwidth,clip]{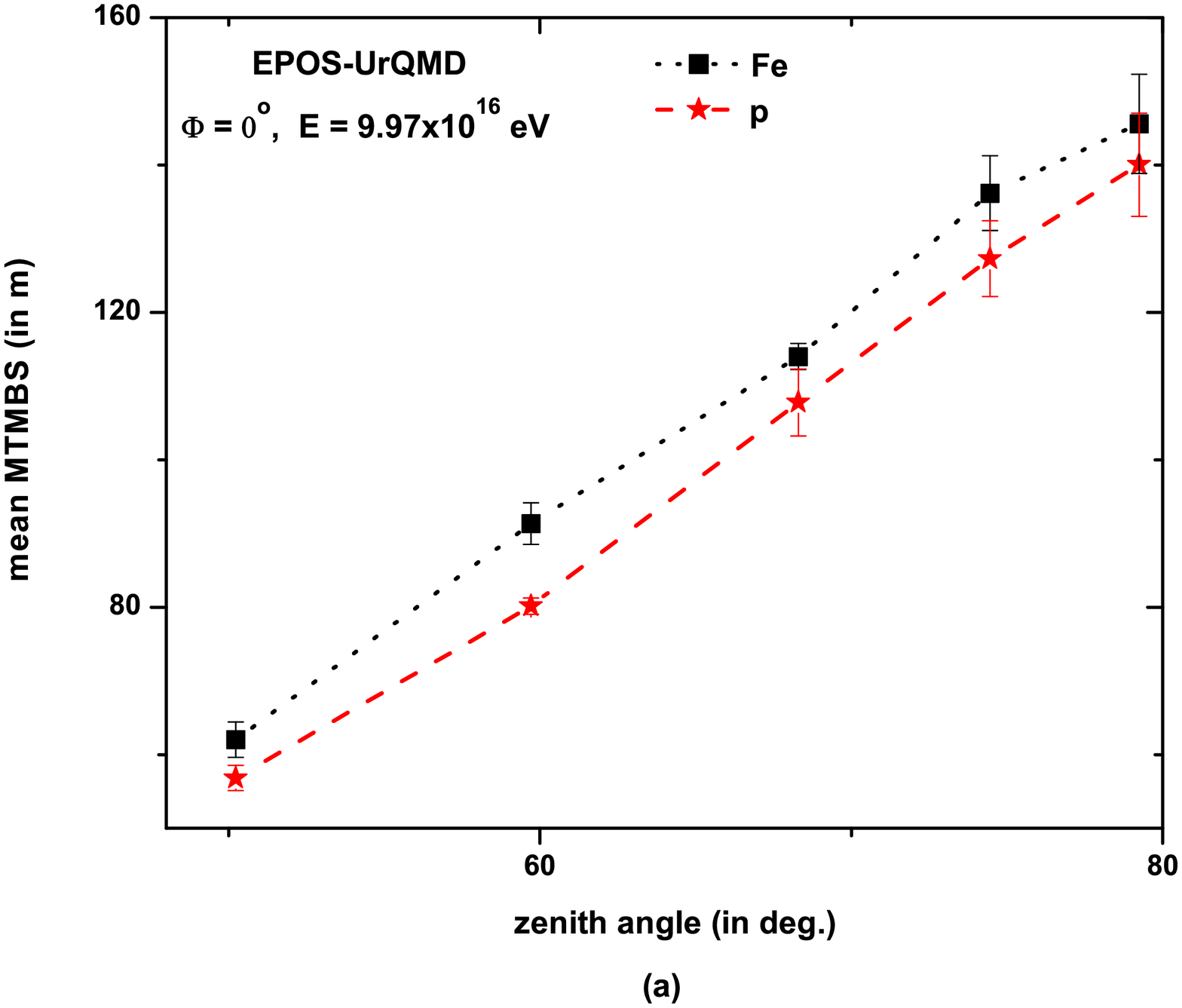} \hfill
\includegraphics[width=0.45\textwidth,clip]{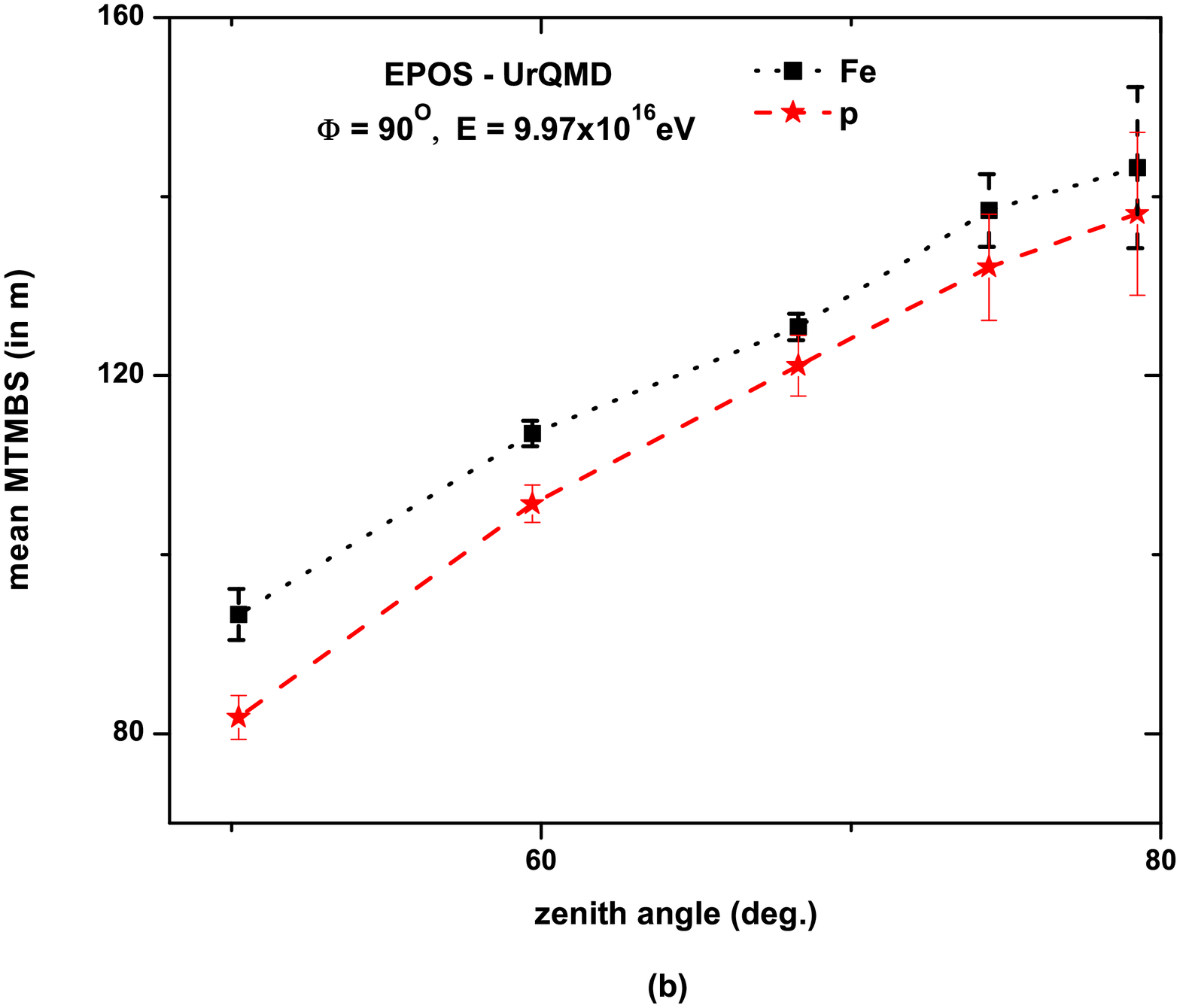} \hfill
\includegraphics[width=0.45\textwidth,clip]{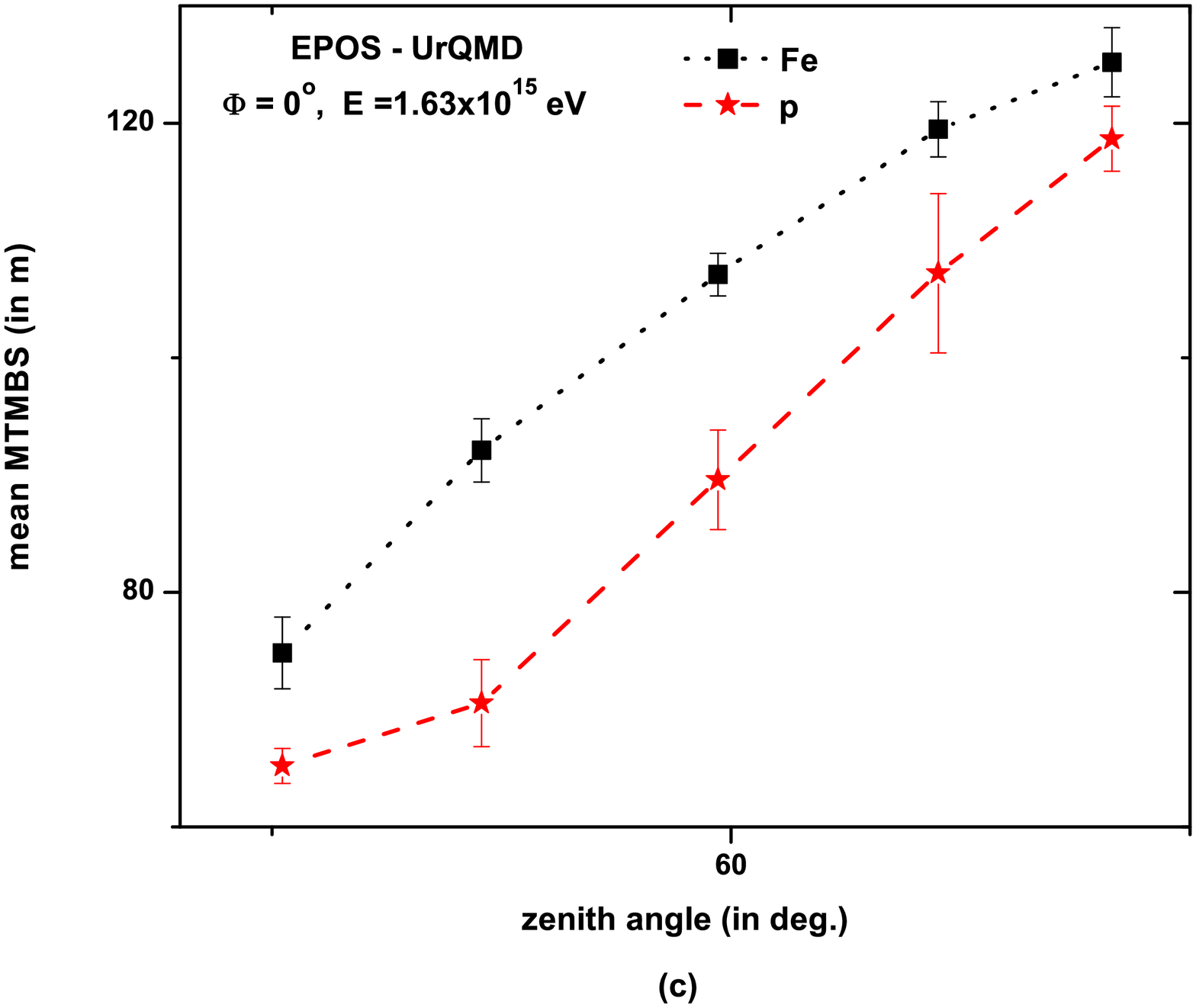} \hfill
\includegraphics[width=0.45\textwidth,clip]{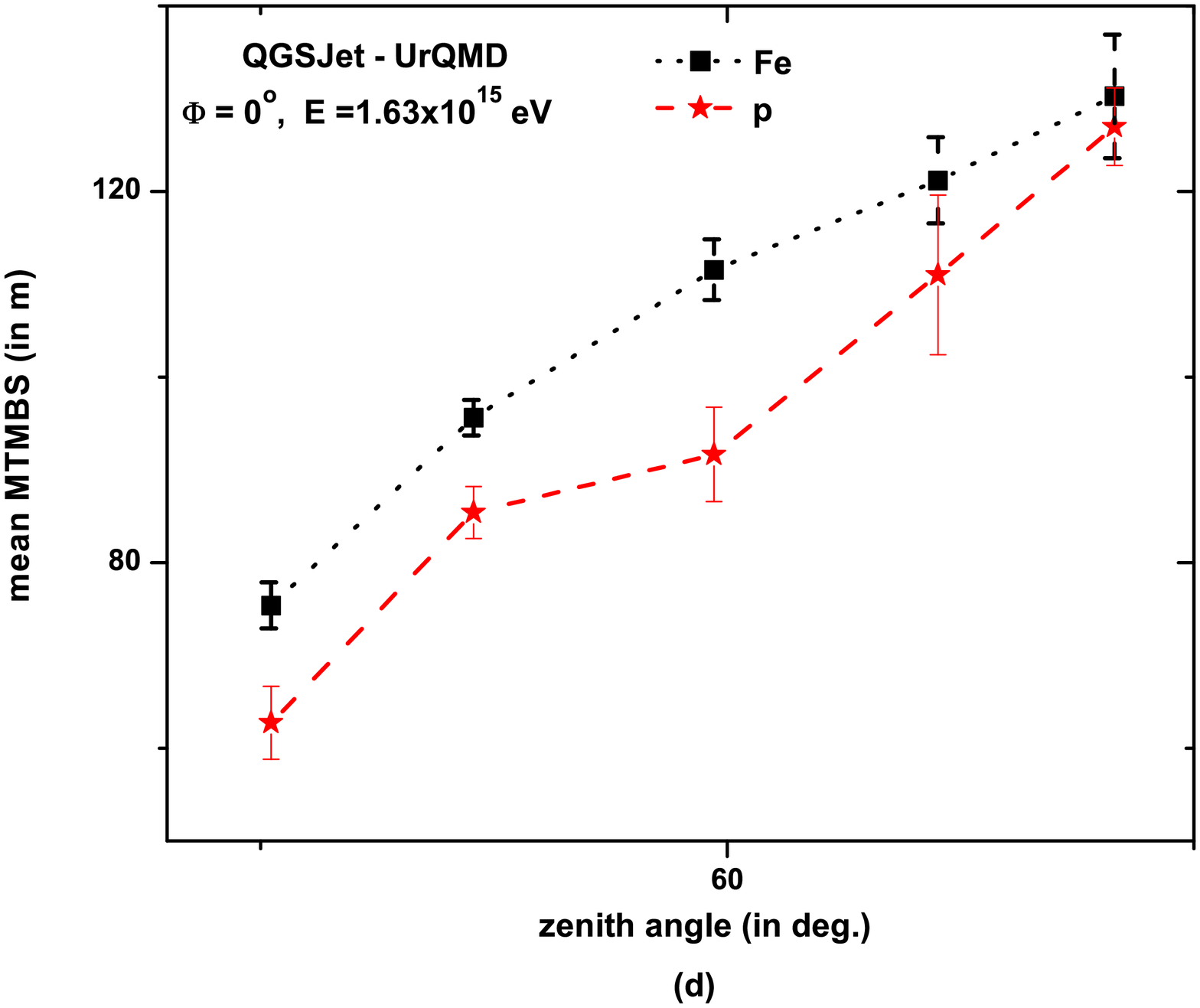} \hfill
\caption{Variation of the mean MTMBS with $\Theta$ and dependence on the high-energy hadronic interaction models. Model dependence is exhibited through comparison of figures c and d. The lines are only a guide for the eye.}
\end{figure}

\begin{figure}
\centering
\includegraphics[width=0.45\textwidth,clip]{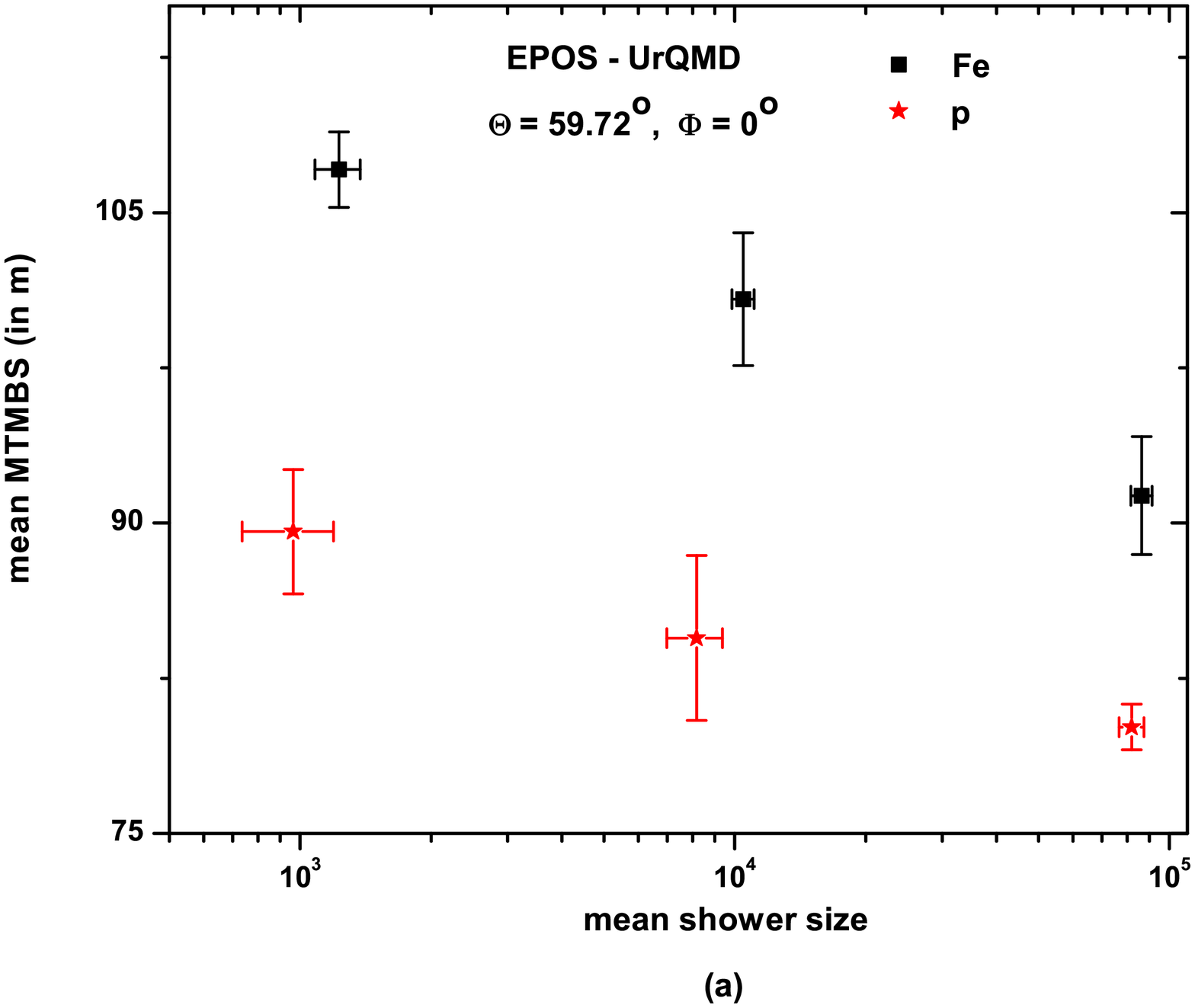} \hfill
\includegraphics[width=0.45\textwidth,clip]{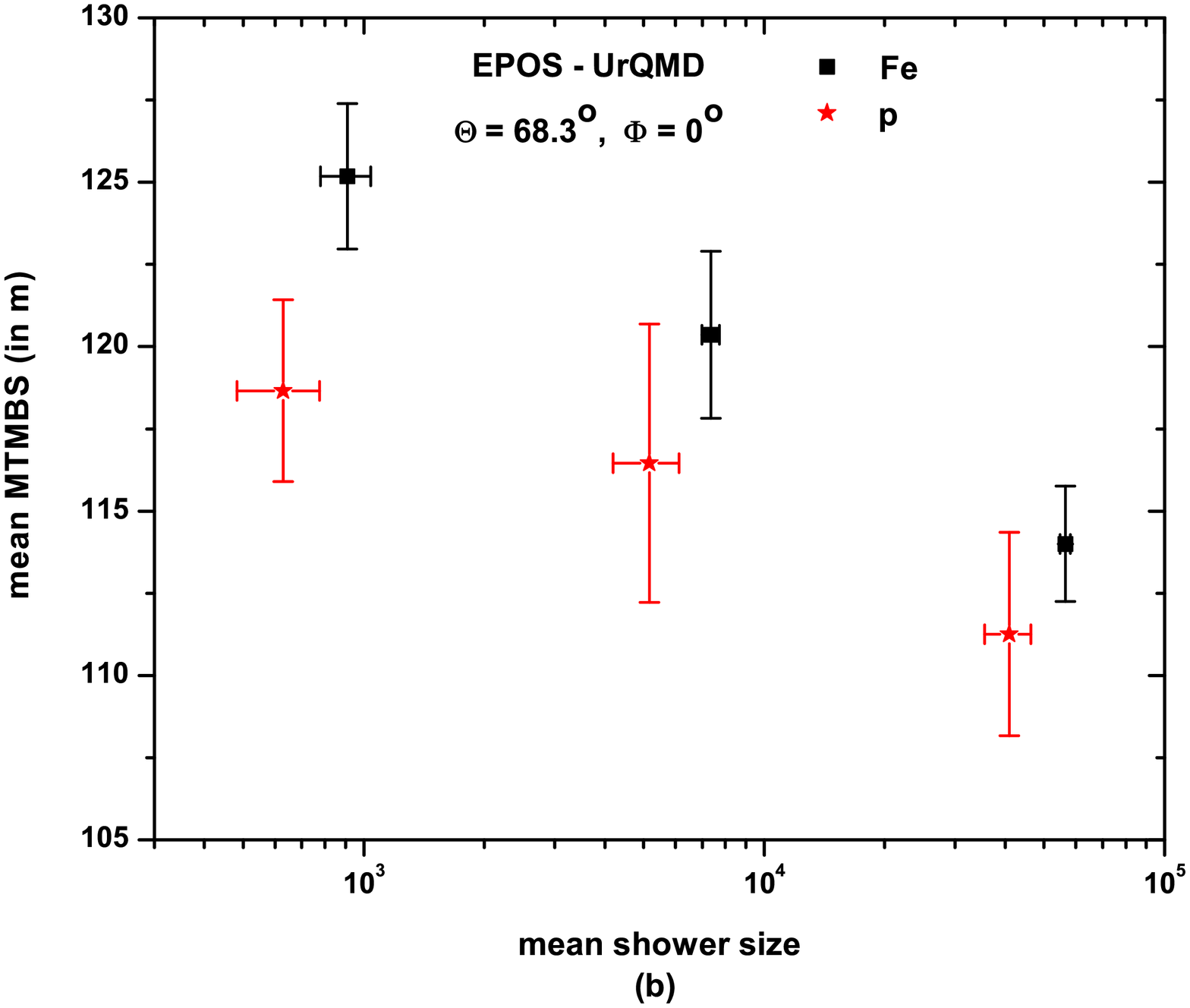} \hfill
\caption{Variation of mean MTMBS with mean shower size for p and Fe initiated showers coming from the North: Fig. a with $\langle \Theta \rangle = 59.72^o$ and Fig. b with $\langle \Theta \rangle = 68.30^o$.}
\end{figure}

To assign a single TMBS to each shower instead of several values corresponding to various positions of the IQS-S over the polar angle between $0^o$ to $180^o$, we have introduced a mean maximum TMBS i.e. MTMBS, which is the average of four TMBS values for four different polar positions of the IQS-S over the polar angle between $\sim 67^o$ to $\sim 112^o$ (between $\sim 247^o$ to $\sim 292^o$ is the region for IQS-S in the opposite side) when $\Phi = 0^o$. For $\Phi = 90^o$, the MTMBS is estimated by taking average of TMBS from the polar angle range between $\sim 22^o$ to $\sim 67^o$ (opposite side: between $\sim 202^o$ to $\sim 247^o$). The parameter MTMBS appears as a suitable mass sensitive parameter for the measurement of CR chemical composition and an experimental approach to estimate it in a ground-based EAS experiment will be discussed in the Sec. VI.

The variation of MTMBS against $\Theta$ is shown in figures 10a and 10b for showers induced by p and Fe, and coming from the North and West directions respectively at $\langle E \rangle = 9.97 \times 10^{16}$ eV. MTMBS values are higher for Fe compared to p primaries in all cases irrespective of $\Theta$, $\Phi$, $E$ and models. According to the figure 10b, MTMBS values for p and Fe showers are approaching closer to one another as the $\Theta$ increases. In figures 10c and 10d, we have presented the effect of high-energy hadronic models on the parameter under consideration. Results obtained from the EPOS look relatively better in comparison with the QGSJeT as far as the composition study of CRs is concerned. Such a behavior of the EPOS validates the generation of little higher $\mu$-s by the model. Errors for MTMBS are evaluated by averaging standard deviations in TMBS values within the polar angle range between $67^o$ and $112^o$ when $\Phi = 0^o$ ( i.e. $\langle \sigma \rangle = (\sum \sigma_i^{2}/4)^{\frac{1}{2}}, i=1,..,4$). For $\Phi = 90^o$, averaging is done using values from the range $22^o - 67^o$.

\subsubsection{Eccentricity parameter of the muon lateral distribution}

The primary CR composition study has also been carried out by the eccentricity parameter ($\epsilon$) of the polar asymmetric distribution of the $\mu$ component under the GF. We have defined the $\epsilon$ parameter as $\epsilon = \sqrt{1 - (\frac{d_{\bot}}{d_{\|}}})^{2}$, where $d_{\|}$ and $d_{\bot}$ denote the core distances of a pair of points possessing same $\mu$ densities in an overall iso-density contour (elliptic) made by $\mu^{+}$ and $\mu^{-}$. One of the densities is estimated along an arbitrary axis which is being formed by the TMBS itself and other, being $\bot^{r}$ to the TMBS axis passing through the EAS core in the shower plane. Actually, $\mu^{+}$ and $\mu^{-}$ in inclined showers under the GF are distributed asymmetrically in the shower plane ($\mu^{+}$ and $\mu^{-}$ distributions also develop elliptic contours covering smaller regions but with some overlapping region). Combination of these smaller contours of $\mu^{+}$ and $\mu^{-}$ will appear as an overall pattern which is equivalent to an iso-density ellipse. Such a pattern of the combined elliptic contour resulting from a pair of smaller iso-density contours is presented by the sketch in the figure 9. The $\epsilon$ factors are evaluated by the above procedure at five $\langle \Theta \rangle$ ranging from $50^{0}$ to $69^{0}$ for p and Fe showers using the QGSJet and EPOS models. We have taken a particular $\Phi = 0^o$ and $\langle E \rangle = 1.63 \times 10^{15}$ eV for the exercise and results are shown in the table III. It reveals from the table that the lateral $\mu$ distribution in the shower plane gets more and more stretched along the axis constituted by the TMBS as $\Theta$ increases. The eccentricity factor correspondingly increases as one moves from lighter to heavier primaries. It has been noticed here that the parameter $\epsilon$ couldn't be estimated by the above procedure for showers with $\Theta \geq 75^{0}$. According to the figure 9 for $\Theta_3$, the stretching of the $\mu$ lateral distribution along the TMBS axis is so high that iso-density concentric ellipse formation couldn't be possible (from iso-density ellipse to stretched out geometric $8$-shaped density pattern). Hence the parameter cannot be useful for very inclined showers to measure the chemical composition of CRs. The parameter $\epsilon$ gets higher values consistently for all $\Theta$ in case of EPOS model compared to the QGSJet. In EPOS, there exists an additional particle production source which induces a larger number of muons from simulations with the model [43]. Hence, an increased number of muons come under the influence of the GF that would result into higher $\epsilon$.      
  
\subsection{Variation of MTMBS with shower size} 

From the experimental standpoint, the correlation of MTMBS with $N_{e}$ has been considered as a basis for extracting information on the nature of primary CRs. The problem of determining the mass composition of CRs is a complicated one because many parameters in tandem are required. From the exploration of the geomagnetic spectroscopy it is found that parameters like TMBS, MTMBS and $\epsilon$ extracted from the asymmetric $\mu^{+}$ and $\mu^{-}$ distributions are found sensitive to the CR mass composition. Practicability of the present approach to determine the CR chemical composition requires a correlation between the observables MTMBS with $N_{e}$ on an average shower with some $\langle \Theta \rangle$ corresponding to either $\Phi = 0^o$ or $90^o$. It might be noted that the present approach on the measurement of primary mass composition with the observable TMBS or MTMBS, a new EAS technique, that is proposed in this work, and therefore no published data/results are available so far for comparison. In figures 11a and 11b, the variation for the mean MTMBS with $\langle N_{e} \rangle$ for p and Fe primaries are shown at $\langle \Theta  \rangle = 59.72^o$ and $68.30^o$ respectively for $\Phi = 0^o$, corresponding to three primary energy ranges $1-3, 8-12, 98-102$ PeV. The figure 11a looks better for composition study compare to figure 11b so far MTMBS versus $\langle N_{e} \rangle$ curve is concerned. Relatively less sensitivity on primary mass at higher $\langle \Theta  \rangle$ might be due to less number of muons surviving against attenuation.        

\begin{figure}
\centering
\includegraphics[width=0.45\textwidth,clip]{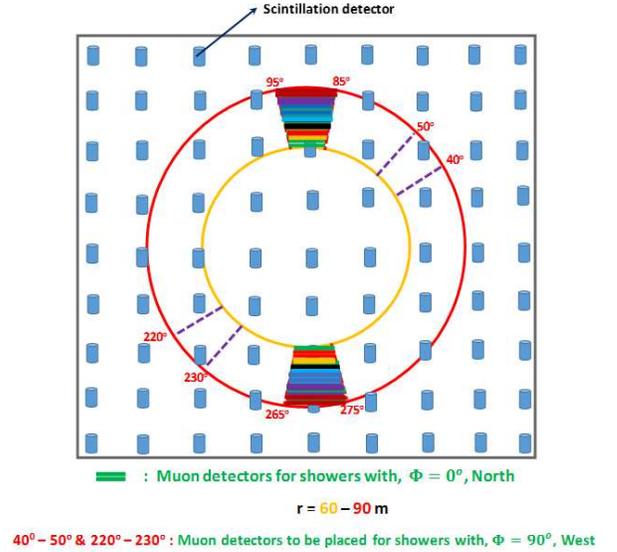} \hfill
\caption{Sketch of a possible array layout containing number of scintillation detectors and two muon detectors for practical realization of the proposed EAS method.}
\end{figure}  
       
\section{Experimental approaches and concluding remarks}
It has been found that the influence of the Earth's magnetic field plays a dominant role for detailed features of $\mu$-s lateral distribution on the shower plane at least for highly inclined showers. The proposed method employs the polar asymmetry in $\mu$-s distribution to determine the primary CR chemical composition at a given observation level for three small energy intervals restricting showers from the North and West directions only. This new EAS method was validated using simulated data samples corresponding to pure proton and iron primaries. Although not done here, there is a scope to extend the present analysis in a much wider primary energy range for the measurement of chemical composition of CRs with numerous features (e.g. any indication of possible change in the chemical composition of PCRs with energy). This can be carried out for other geomagnetic positions as well to correlate some geomagnetic properties with high-energy astro-particle phenomena. In order to present important results at fixed shower size and muon size under different geomagnetic intensities, a high-speed computation with adequate storage capacity has to be ensured.
 
Our analysis concerning the effects of the GF on $\mu^{+}$ and $\mu^{-}$ particles of inclined EASs reveals several interesting features such as polar asymmetries, sectoral $\mu$ abundances, elliptic footprints of $\mu$ distributions, and amplitude of fluctuations between proton and iron induced showers. Taking a small sample of data such effects are also found to persist with comparable magnitude (not shown here) if we replace the UrQMD code in the simulation by the Fluka [44] in the treatment of low energy hadron collisions.

The present method might help to design EAS arrays with induction of $\mu$ detectors to derive more information about the primary CRs employing the trio {\bf EAS direction} ($\Theta$ and $\Phi$), {\bf EAS core} ($x_0$, $y_0$) and {\bf geomagnetic field}. For implementing the present technique in a possible experimental set up, resolution of the EAS array for selected $\Theta$ ranges corresponding to some fixed or $\Phi$-range at a given geomagnetic position should be a prerequisite. This is usually realized by performing MC simulations of EASs, convoluted with the EAS array structure and detector responses. We know that at highly inclined showers (present work) the angular resolution of an array generally deteriorates with increasing $\Theta$. A reasonable array resolution should be provided for implementing the present method from the employment of MC showers. A certain triggering level has to be set by some central triggering detectors in the array using MC showers. 

For the present purpose, the EAS experimental set up firstly should have a close-packed detector array ($200$ m x $200$ m like KASCADE) for determining global parameters of an EAS (here, $\Theta$, $\Phi$ and $x_0$, $y_0$ etc.). It should be mentioned that for a closely packed array, estimation for the EAS core location by simple weight averaging of particle ($e^{\pm}$) densities from discrete detector locations does not differ appreciably from the core obtained from the shower reconstruction. Secondly, estimation of the main parameter in the work i.e. the MTMBS, one has to incorporate a pair of $\mu$ detectors with sizes at least covering the present regions around $90^o$ and $270^o$ in a possible geometrical layout shown in the figure 12 for $\Phi = 0^o$. Two uncovered sectors in the figure 12 within a pair of dotted lines for each, in opposite sides within the polar regions of $40^o - 50^o$ and $220^o - 230^o$ (possible locations for $\mu$ detectors for EASs coming from the West). The sizes for $\mu$ detectors proposed here are having the least size from the point of view of their production cost. As a result, the measurement of the MTMBS may involve slightly higher uncertainty. As we know that a $\mu$ detector consists of number of proportional counters having some energy threshold (here, $\geq10^2$ GeV). In the figure 12, the green patch shows $\mu$ detectors layer which is under a shielding absorber containing several layers (concrete, soils, lead or iron sheets etc. shown by different color patches) to set the threshold energy at $10^2$ GeV for $\mu$-s. It would be an interesting task to apply such a method using observed EAS data if the KASCADE or GRAPES - III experiments runs with the facility of concurrent $\mu$ measurements ($\mu$ tracking detectors (MTD) [45]) after restructuring the present $\mu$ detecting systems. A new experiment consisting of underground installations of muon detectors and an EAS array above the large underground would be appropriate for the study.

There are some recent proposals of studying positive and negative $\mu$-s separately in individual EAS event. In fact, a few ongoing experiments, such as the WILLI detector [26] or the Okayama University, Japan EAS installation, have the capability to extract charge information of high energy $\mu$-s but these experiments do not have large $\mu$ detection area, which is needed to extract information about the nature of primaries from the study of geomagnetic influence on EAS $\mu$-s. If in future, these experiments are extended in order to cover larger detection area, or new installation of large $\mu$ detection area with capability of charge identification are come up, the present proposal can be exploited to extract the nature of primary particles.

The main sources of error are the fluctuations in $\mu$ numbers from shower to shower, non-uniform behavior of the local geomagnetic field, and the uncertainties in arc, and radial distance estimations by $X, Y$ values of each $\mu$ within a polar region of a selected ISQ-S. From the cost-productive viewpoint, a slim ISQ-S (figure 6 or 12) with truncated radial bin ($60 - 90$ m) covering a pair of $\mu$ detectors in opposite sides has been proposed in a real experiment. But, in such a situation a slightly higher uncertainty in the parameter estimation may result.  
 
\section*{Acknowledgments}
RKD thanks the SERB, Department of Science and Technology (Govt. of India) for financial support under the grant no. EMR/2015/001390.

\end{document}